\documentclass[prb,floatfix,preprint]{revtex4}
\usepackage[final]{graphicx}
\usepackage[usenames]{color}
\usepackage{afterpage}

\begin{document}

\title{Static and Dynamical  Susceptibility of LaO$_{1-x}$F$_x$FeAs}
\author{ M. Monni$^1$, F. Bernardini$^1$,  G. Profeta$^2$,  
A. Sanna$^{3,4,1,5}$, S. Sharma$^{3,5}$, J. K. Dewhurst$^{3,4,5}$, C. Bersier$^{3,4,5}$,  A. Continenza$^2$, 
E.K.U. Gross$^{3,5}$, S. Massidda$^1$}
\affiliation{1 SLACS-INFM/CNR,  and Dipartimento di Scienze Fisiche, 
Universit\`a degli Studi di Cagliari, I-09042 Monserrato (CA), Italy}
\affiliation{2 CNISM - Dipartimento di Fisica,
Universit\`a degli Studi dell'Aquila, Via Vetoio 10,
I-67010 Coppito (L'Aquila) Italy}
\affiliation{3 Institut f{\"u}r Theoretische Physik, Freie Universit{\"a}t
Berlin, Arnimallee 14, D-14195 Berlin, Germany}
\affiliation{4  Fritz Haber Institute of the Max Planck Society, Faradayweg 4-6, 
D-14195 Berlin, Germany.}
\affiliation{5 Max-Planck-Institut f\"ur Mikrostrukturphysik, Weinberg 2, 
D-06120 Halle, Germany.}
%\email{sandro.massidda@dsf.unica.it}
%\thanks{}

\begin{abstract}

The mechanism of superconductivity and magnetism and their possible interplay 
have recently been under debate in pnictides. A likely pairing mechanism 
includes an important role of spin fluctuations and can be expressed in terms 
of the magnetic susceptibility $\chi$.  The latter is therefore a key quantity 
in the determination of both the magnetic properties of the system in the 
normal state, and of the contribution of spin fluctuations to the pairing potential.  A basic ingredient to obtain $\chi$ is the 
independent-electron susceptibility $\chi_{0}$.
Using LaO$_{1-x}$F$_{x}$FeAs as a prototype material,  in this report we 
present a detailed \emph{ab-initio} study of $\chi_{0}({\bf q},\omega)$, 
as a function of doping and of the internal atomic positions.
The resulting static $\chi_{0}({\bf q},0)$ is consistent with both the observed
 \emph{M}-point related magnetic stripe phase in the parent compound, and with 
the existence of incommensurate magnetic structures predicted by 
\emph{ab-initio} calculations upon doping.
\end{abstract}
\maketitle

\section{introduction}

Iron pnictides\cite{Kamihara} represent a challenge in the field of superconductivity, both
from the experimental and theoretical point of view\cite{expreview,mazinreview}.
They are magnetic metals that upon electron (\emph{e}) or hole (\emph{h}) doping
transform into high temperature superconductors. The prototype of this new family of superconductors
is LaOFeAs\cite{Kamihara}, which superconducts at 26K 
upon partial substitution ($\approx 14$\%) of O with F.  Several other superconducting
pnictides have since been discovered, with a record critical temperature 
up to $T_{c}\approx 55 $ K\cite{Sm}. 
%%%%%%%%%%%%%%%%%%% High light spin fluct in liteature.
The parent undoped compound for these superconductors are anti-ferromagnetic (AFM)
semi-metals, with several competing magnetic structures lying within  a few tens of 
a meV\cite{Dong08,Singh08,DMFT,Mazin,Xu08,Cao08,sharma,yaresko}. 
This is an ideal situation for the presence of magnetic instabilities, with a 
strong possibility of spin fluctuations playing an important role. In fact this 
is confirmed by experiments, which show that the suppression of the magnetic 
instability is a necessary step to obtain superconductivity \cite{norman,wakimoto}.

%%%%%%%%%%%%%%%%%%%  Differences between cuprates and FeAs: correlations
This proximity  to an AFM instability leads one to draw 
parallels between pnictides and cuprates.  
Despite several similarities these two families of superconductors show important 
differences. One of the most important differences is related 
to electronic correlations; cuprates are well known to be strongly correlated 
materials, where a treatment of on site Coulomb interactions 
is essential to get the correct normal ground state. On the 
other hand in the case of pnictides it was concluded by Anisimov et al. \cite{anisimovsmallU} 
that the electronic structure of these materials is consistent with a small 
value of $U$ (within LSDA+$U$ scheme) indicating that strong correlations may not be 
essential to describe them. In fact, unlike in cuprates, standard 
local/semi local functionals within density functional theory (DFT) can describe 
the occurrence of magnetism in pnictides, although the agreement between the 
calculated and the experimental magnetic moment and its dependence upon the choice
of functional and atomic positions has been extensively discussed\cite{mazinreview, sharma}.

%%%%%%%%%%%%%%%%%%%%%%% superconducting
%%%%%%%%%%%%%%%%%%%%%%%%%%%%%%%%%%%%%%%%%%%%% Phonons and electronic mechanism
There is not a general consensus about the mechanism leading to superconductivity 
in these materials.  It seems clear, however, that the electron-phonon
interaction alone (without involvement of spins) is too weak\cite{Boeri} to
produce such a high $T_{c}$, at least within the standard Migdal-Eliashberg theory. 
On the other hand explicit involvement of the spin degree of freedom leads to an 
enhancement of electron-phonon coupling\cite{Yildirim,Yndurain}; therefore the role 
of phonons has not been completely ruled out.

However, the presence of a weak electron phonon coupling at least in the classic 
sense lead to the suggestion of several alternative mechanisms which are essentially 
electronic in nature. Among these alternatives one of the most 
prominent is the mechanism suggested by
Mazin et al\cite{Mazin}; inspired by the peculiar shape of
the Fermi Surface (FS)\cite{ARPES} they suggested an $s_{\pm}$ superconducting order parameter,
having different sign on the \emph{h} and \emph{e}-like sheets forming the FS. In this way superconductivity can be driven by a strong repulsive inter-band 
interaction (like spin fluctuations). The role of spin fluctuations in the Cooper pairing was
proposed several years ago by Berk and Schrieffer\cite{Berk66}; this mechanism
requires a non-conventional
(\textsl{i.e.} not a simple $s$-wave) superconducting order parameter.
The effective Hamiltonian of a system close to a magnetic instability
"pairs" states at $\mathbf{k}$ and $\mathbf{k+q}$ by an effective interaction
matrix element, ultimately related to the $\mathbf{q}$-vector and
frequency dependent susceptibility $\chi(\mathbf{q},\omega)$. Both the
$\mathbf{q}$-vector and frequency dependence of $\chi$ are
important in determining the symmetry and possibly the anisotropy of the
superconducting gap on the FS.
Experiments, on the other hand, cannot make a definite statement on the symmetry
of the order parameter\cite{expreview,mazinreview} making it difficult to
validate/invalidate this scenario.

It is clear that the above mentioned electronic mechanisms rely crucially 
on the detailed knowledge of the susceptibility, $\chi(\bf{q},\omega)$.
Several calculations of the static $\chi$ have been reported on pnictides. 
The first \emph{ab-initio} calculation of the $\chi({\bf{q}},0)$ did 
not include matrix elements among Bloch states\cite{Mazin}. Later 
calculations\cite{scalap1,kuroki,eremin} of $\chi_0$ were mostly based on the
tight-binding fits of the minimal bands manifold around  
$E_{F}$, without inclusion of higher energy inter-band transitions. 
This approach suffers from the fact that, even though the main structure of the static
$\chi(\mathbf{q},0)$ can be determined by low energy transitions (without 
including many empty bands) no definite conclusions can be 
reached about the structure of $\chi(\bf{q},\omega)$. Hence an 
\emph{ab-initio} investigation of the full dynamical susceptibility, including 
a proper account of the 
matrix elements involving Bloch states is highly desirable and is still lacking.
To fill this gap, in this article we present an \emph{ab-initio} determination of 
the static and dynamical
independent-electron susceptibility of LaOFeAs, based on electronic structure 
calculations performed within DFT. The results will be
a first step towards an understanding of the material properties and of the 
possible contribution of spin fluctuations to the superconducting pairing.  

The rest of the paper is organized as follows: Section \ref{meth} discusses
the methodology and computational details. Section \ref{resul} contains
the results. In particular, first the static susceptibility
$\chi_{0}(\mathbf{q},0)$ is discussed and then the results for the real 
and imaginary parts of $\chi_{0}$ as a function of frequency are presented.
Finally the conclusions are given in Section \ref{concl}. 

%%%%%%%%%%%%%%%%%%%%%%%%%%%%%%%%%%%%%%%%%%%%%%%%%%%%%%%%%%%%%%%%%%%%%%%%%%%%%%%%
\section{method}
\label{meth}
The independent electron susceptibility is defined as:\begin{equation}
\chi_{0}\left(\mathbf{q\mathrm{,}G\mathrm{,}G'\mathrm{,\omega}}\right)=
\sum_{nn'k}\frac{f_{nk}-f_{n'k+q}}{\epsilon_{nk}-\epsilon_{n'k+q}+
\hbar(\omega+i\eta)}\left\langle n'\mathbf{k}+\mathbf{q}
\left|\mathbf{e^{i(\mathbf{q}+\mathbf{G})\cdot r}}
\right|n\mathbf{k}\right\rangle\left\langle n\mathbf{k}\left|
\mathbf{e^{-i(\mathbf{q}+\mathbf{G'})\cdot r}}\right|n'\mathbf{k}+
\mathbf{q}\right\rangle \label{eq:1}\end{equation}
where $\epsilon_{nk}$ and $f_{nk}$ are the one-electron energies and the 
corresponding Fermi functions, and $\mathbf{G}$,$\mathbf{G'}$ are reciprocal 
lattice vectors.  In this work we accurately
compute $\chi_{0}$ by performing the summations in Eq.~(\ref{eq:1})
using a random sampling over the Brillouin zone (BZ).
We used $\sim$3000 independent $\mathbf{k}-$points per band, chosen according to a stochastic algorithm which
 accumulates them around the FS for bands crossing $E_F$. The final 
results are obtained by averaging over 40 runs each containing completely 
independent {\bf k}-point set. This procedure nearly completely eliminates the
numerical noise, and shows a good convergence both in terms of number of 
independent runs and number of {\bf k}-points within a single run.
We have included 65 bands in order to ensure convergence with respect to the 
number of empty bands. 

The energy bands and the matrix elements for LaOFeAs have been calculated, 
within the (spin-independent) local density approximation (LDA)\cite{vBH} 
to the exchange-correlation functional. Our calculations have been done using 
the Full-potential Linearized Augmented Plane Wave (FPLAPW) method. This choice  
is necessary because of the extreme sensitivity of the electronic structure 
(in particular, of the bands close to $E_{F}$) to the method\cite{Mazin_pseudo}.  
No numerical approximation has been made in the evaluation of matrix elements in Eq.~(\ref{eq:1}).
Given the huge debate about the dependence of the results on the position of the 
As atom in the unit-cell\cite{Mazin_pseudo,Pickett}, in the present work we show results both for the 
theoretically optimized ( $z_{\rm{As}}=0.638$) and experimental\cite{huang} ($z_{\rm{As}}=0.6513$)
position of the As atom.

%%%%%%%%%%%%%%
% Results   %%
%%%%%%%%%%%%%%

\section{results and discussion}
\label{resul}

\begin{figure}
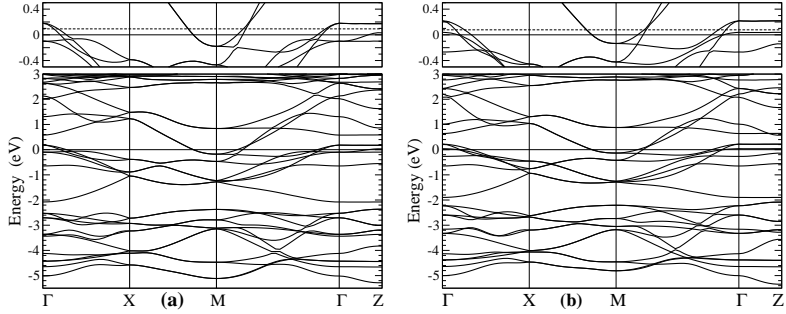

\includegraphics[scale=0.22]{fig_1a.eps}
\includegraphics[scale=0.22]{fig_1b.eps}
\caption{Energy bands of LaOFeAs, for the theoretically optimized (left) and 
experimental position (right) of the As atom, along various symmetry 
directions in the Brillouin zone. The upper panel highlights the region around the
Fermi level. Dashed line in the upper panel indicates the position of the Fermi level for the 14\% doped LaOFeAs.}
\label{fig:bande}
\end{figure}

We first look at the band structure and the FS for the parent 
and the doped LaOFeAs. 
The ground state calculations are performed with the tetragonal unit cell, 
containing two Fe atoms, at the experimental lattice constants 
$a=4.03 $\AA \, and $c/a=2.166$. 
 The energy bands of LaOFeAs calculated using both $z_{\rm{As}}$ 
are shown in Fig. \ref{fig:bande}. Four bands cross the Fermi level $E_{F}$, having 
predominantly Fe $d$ character moderately hybridized with As $p$. These band 
structures are in agreement with previous results\cite{Boeri,Mazin_pseudo}. 

\begin{figure}
\includegraphics[scale=0.17]{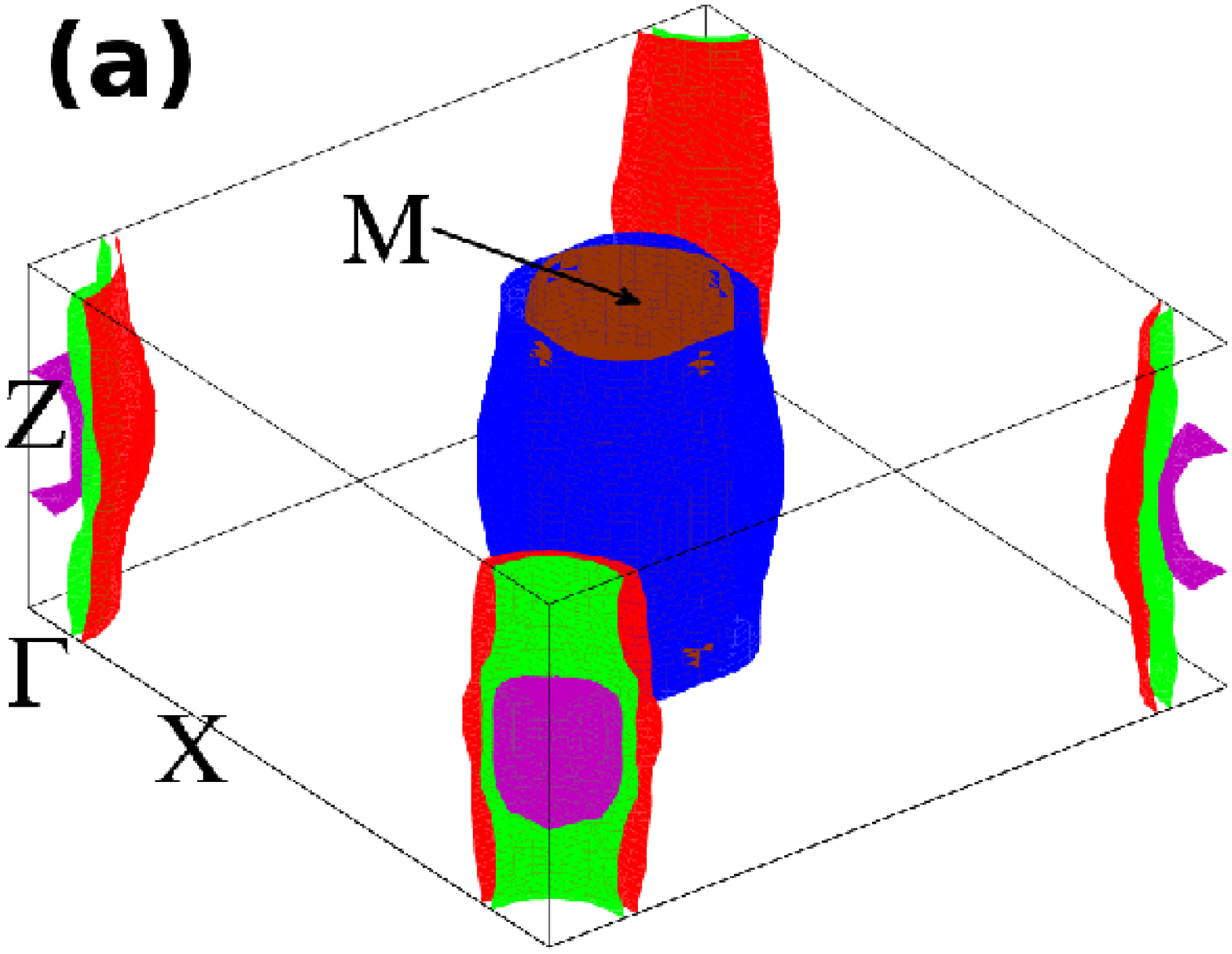}
\includegraphics[scale=0.17]{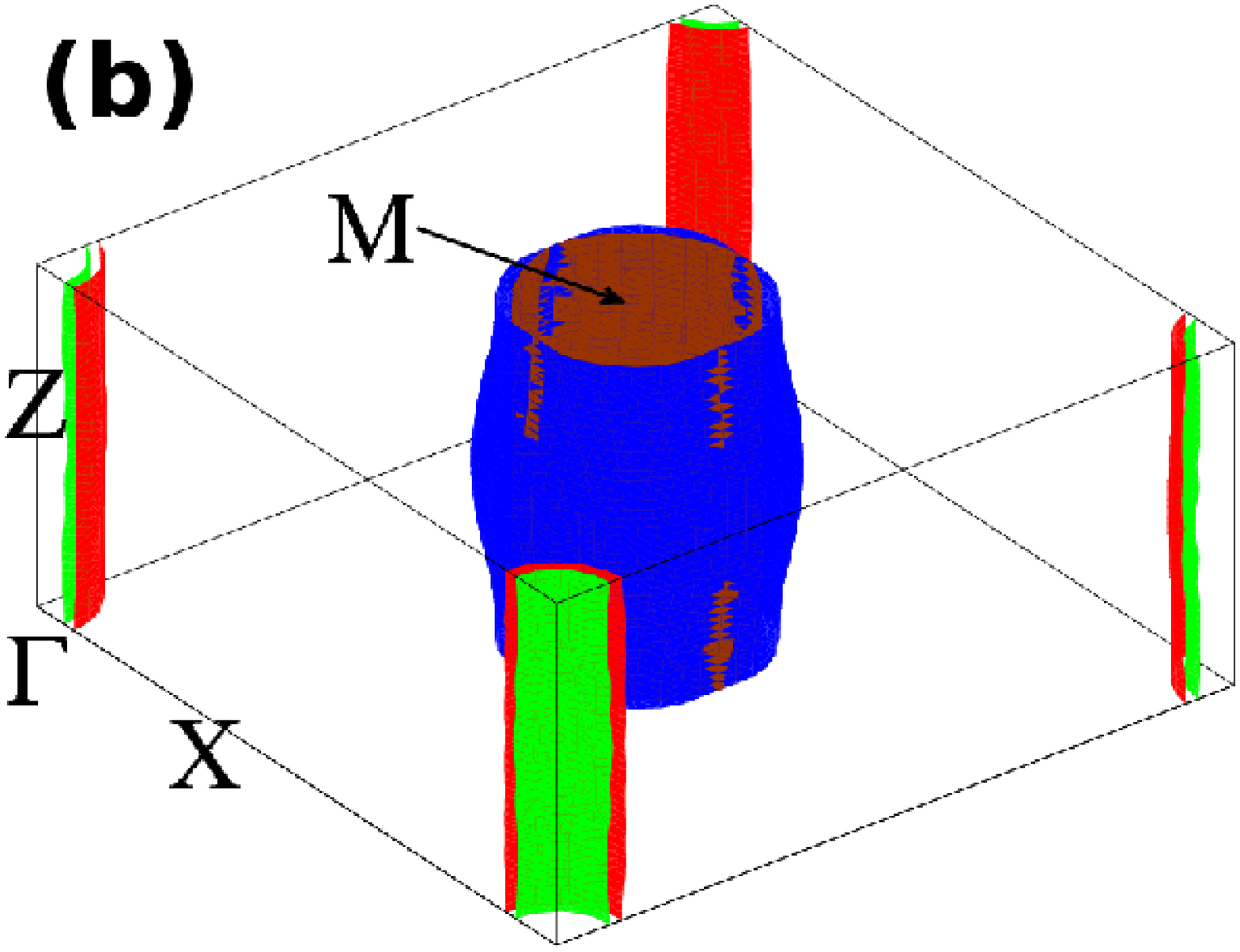}\linebreak
\includegraphics[scale=0.17]{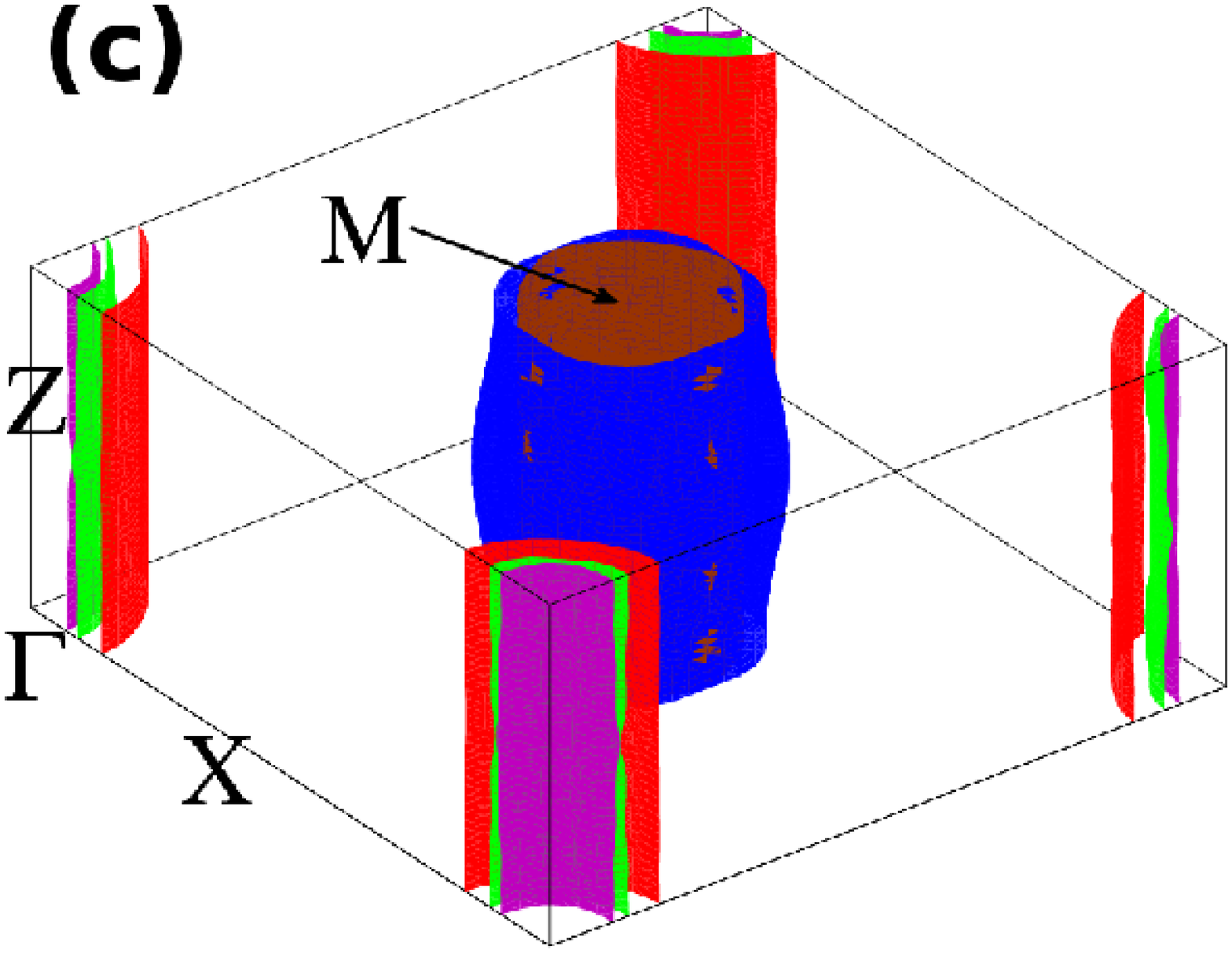}
\includegraphics[scale=0.17]{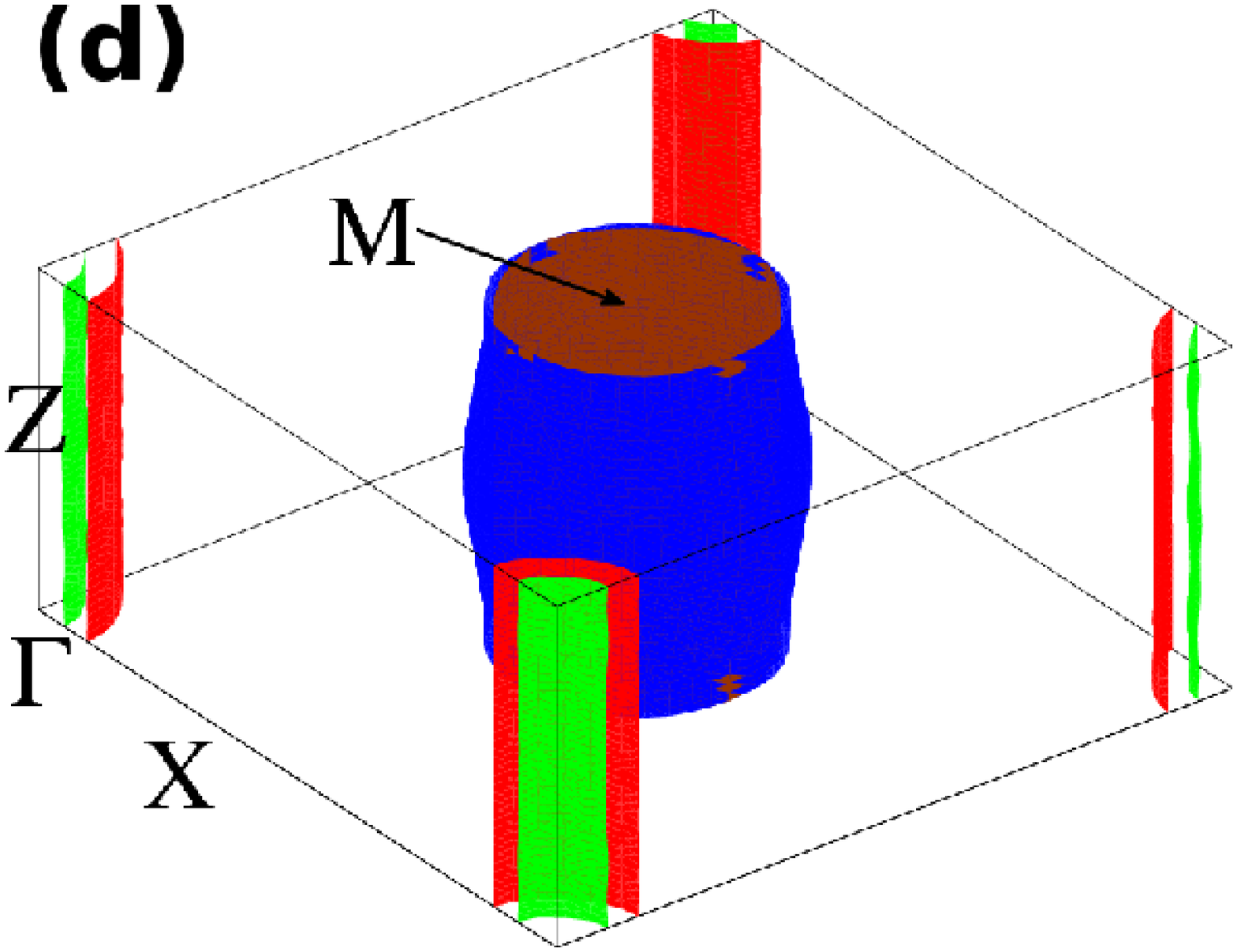}
\caption{Fermi surface of LaO$_{1-x}$F$_{x}$FeAs.
Top-panel: at theoretically optimized position of the As atom with 
(a) $x=0$ and (b) $x=0.14$.
Bottom-panel: at experimental position of the As atom with 
(c) $x=0$ and (d) $x=0.14$. }
\label{fig:FS}
\end{figure}

The corresponding FS are given in Fig. \ref{fig:FS}. The FS 
calculated with optimized and experimental $z_{\rm{As}}$ for the undoped compound
are quite similar with two \emph{e}-like cylinders around the $M$ point and two 
\emph{h}-like cylinders warped around the $\Gamma$ point of the BZ. However, the two 
FS differ when it comes to the third \emph{h}-like manifold which appears as a 3D  
structure on the use of optimized As positions and as a 2D cylinder when the
experimental atomic positions are used. 

Since the parent compound becomes superconducting on electron doping, it is 
important to look at the change in the electronic structure as a function of the Fluorine content.
The doping is treated by the means of rigid band model (RBM) which is a reasonable approximation for LaOFeAs\cite{computational_details}. 
As for the crystal structure, it is kept fixed to the
undoped values even for the doped case. This choice is justified for the material
under investigation; it was shown by Mazin et al. \cite{Mazin_pseudo} that
the lattice parameter and atomic positions are not sensitive to the doping. 
The  FS  for the 14\% doped compound is shown in Fig. \ref{fig:FS}.
Even though the undoped FS calculated using optimized and experimental 
$z_{\rm{As}}$ are substantially different, the doped FS for the two cases are 
fairly similar. The \emph{h}-like FS sheet around $\Gamma$ point for $x=0$ quickly 
disappears on electron doping in both cases ($z_{\rm{As}}$ optimized 
and experimental) with only 2D tubular structures around the $\Gamma-Z$ 
(\emph{h}-like) and the $M-A$ (\emph{e}-like) lines surviving. 
Most noticeably the nesting between \emph{h}-like and \emph{e}-like FS 
present in the undoped material becomes less prominent on doping due to the
increase in the asymmetry between the \emph{e}-like and the \emph{h}-like cylinders.

\subsection{Static response function}
\begin{figure}
\includegraphics[scale=0.2]{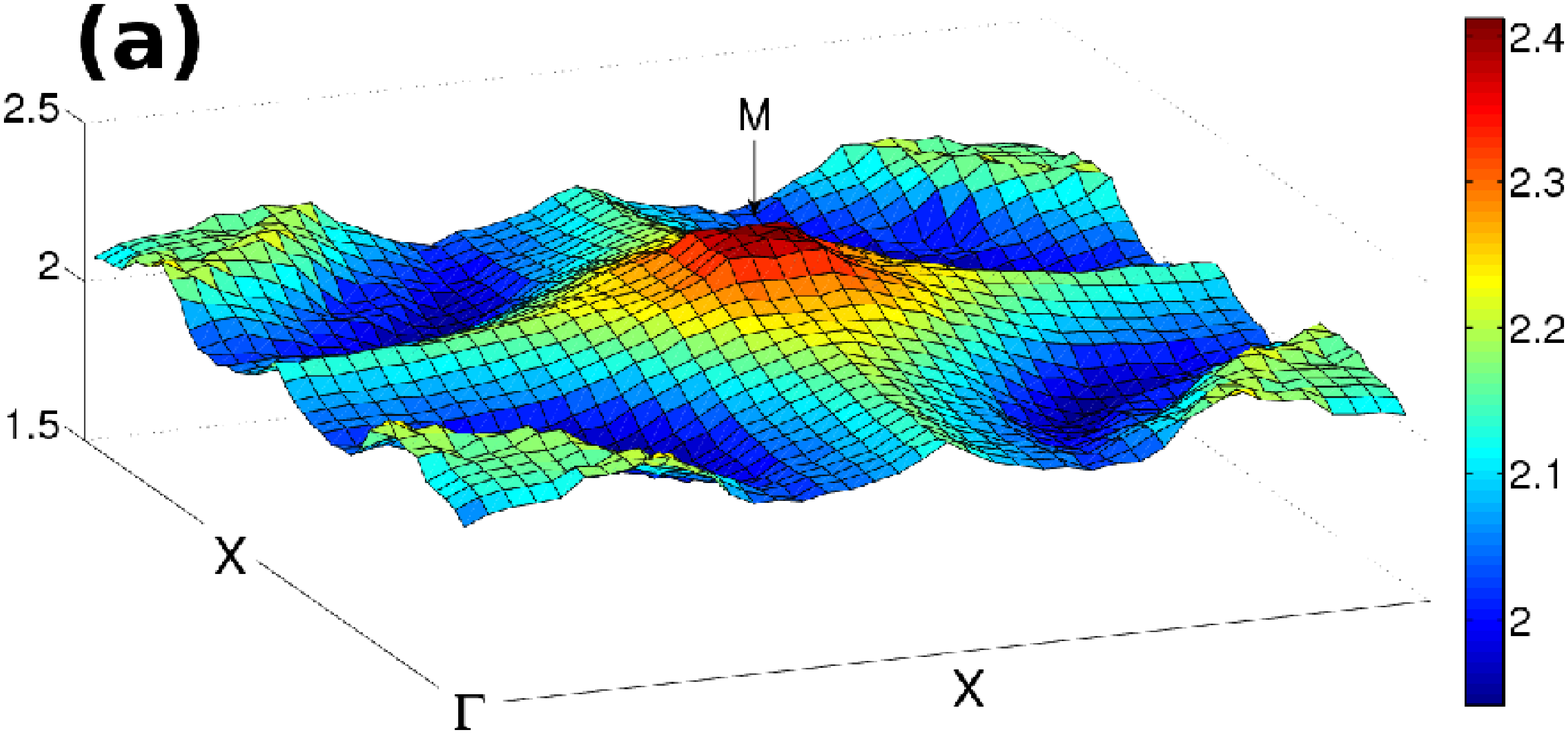}\includegraphics[scale=0.2]{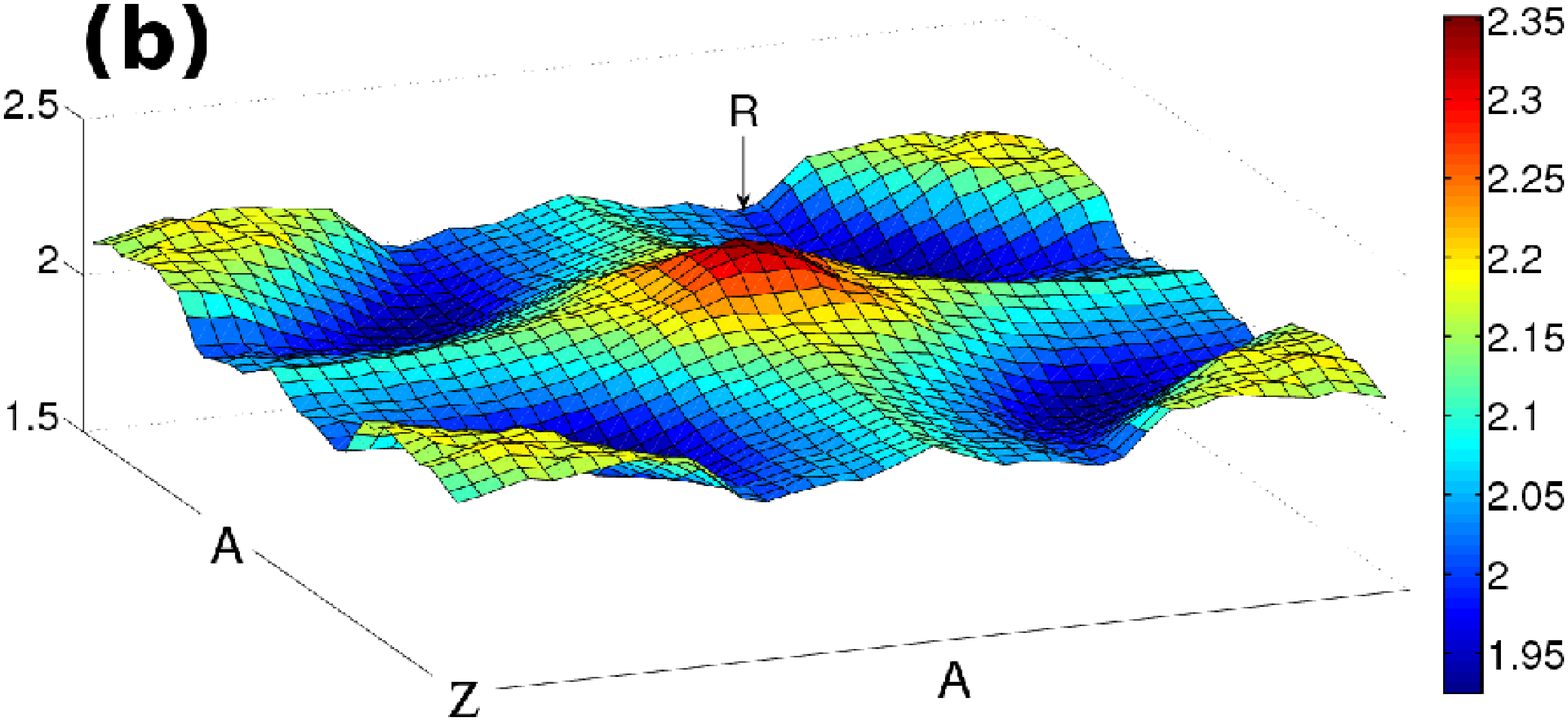}
\caption{Real part of $\chi(\mathbf{q\mathrm{,0)}}$ for undoped LaOFeAs 
(a) on the basal plane (i.e. $q_{z}=0$) and 
(b) on the top plane (i.e. $q_{z}=\frac{\pi}{c}$).}
\label{fig:rechix0}
\end{figure}
\begin{figure}
\includegraphics[scale=0.2]{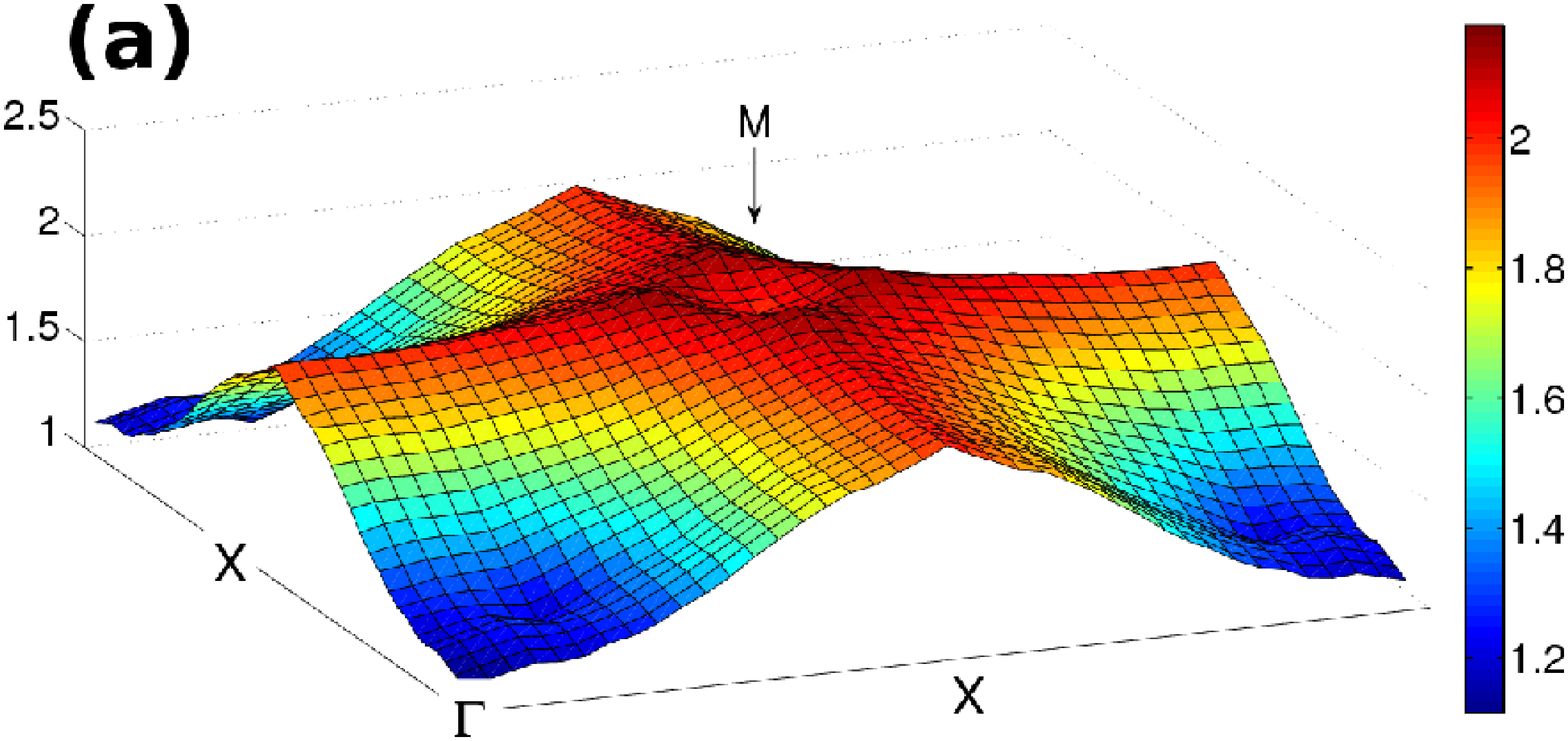}\includegraphics[scale=0.2]{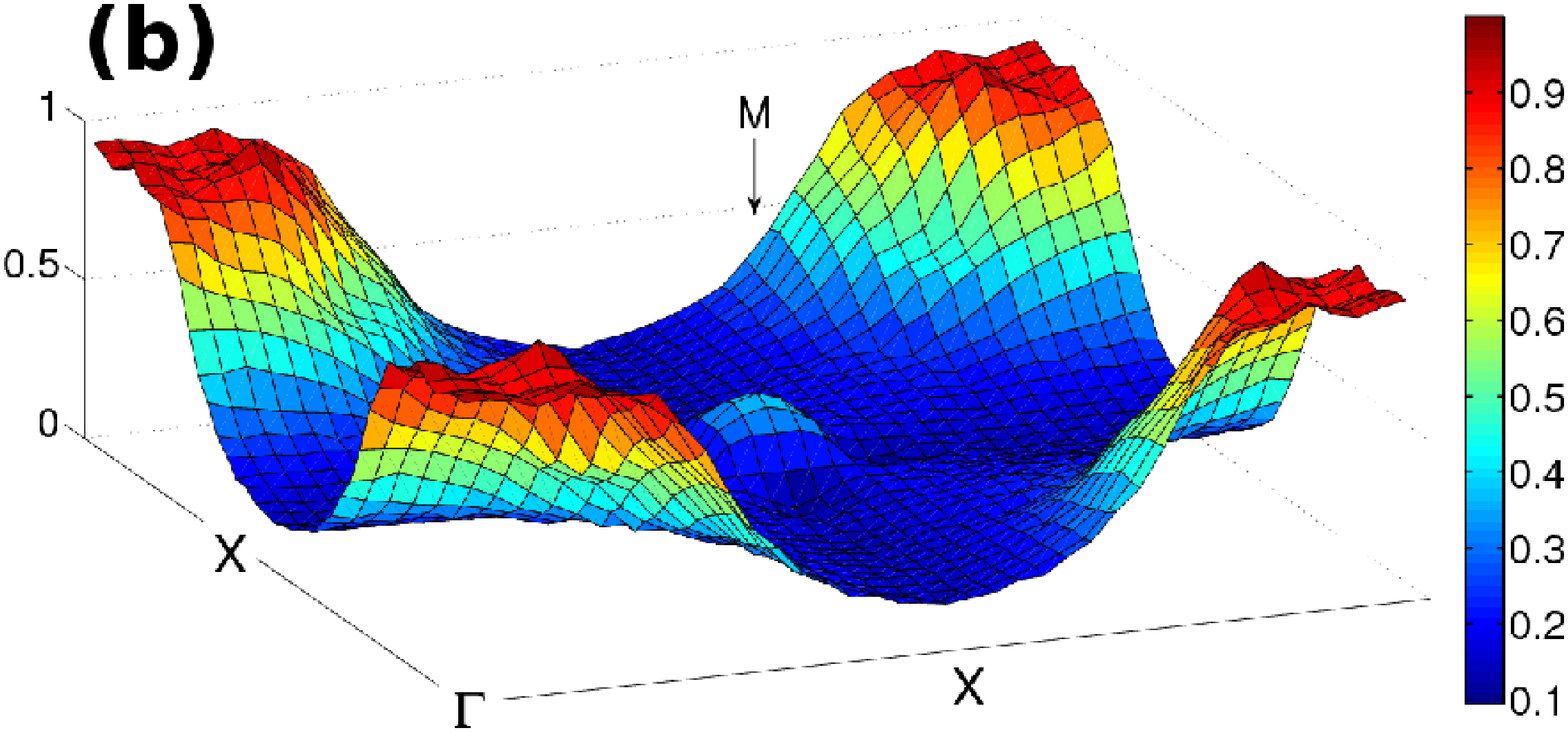}
\caption{ (a) Real part of $\chi(\mathbf{q\mathrm{,0)}}$ for 14\% doped LaOFeAs 
and (b) $\Delta\chi_{0}$: the difference between the real part of $\chi$ for the
doped and undoped LaOFeAs. The results are plotted on the basal plane.}
\label{fig:rechix014}
\end{figure}

%
%-------------------------------------------------------------------------------------------------------
In order to elucidate how this change in FS upon doping manifests itself in the
response of the material, in 
Fig. \ref{fig:rechix0} and  \ref{fig:rechix014} are shown the 
response function $\chi_{0}(\mathbf{q\mathrm{,0)}}$  for LaO$_{1-x}$F$_{x}$FeAs at $x=0$ and $x=0.14$. 
Looking first at the undoped case, $x=0$ (Fig. \ref{fig:rechix0}(a)), 
two main features are clearly visible; first is the presence of  a wide flat 
structure around the $\Gamma$ and a relatively sharper peak around $M$ point and 
second is the flatness of the response function in the $c$-axis (tetragonal axis). 
As expected, due to the 2D nature of this material similar features (structures 
around $Z$ and $R$ points in otherwise flat response function) are also visible 
in the top plane (Fig. \ref{fig:rechix0}(b)). 
These main features of the response function can be explained
on the basis of the electronic structure of LaOFeAs and by comparison with the
non-interacting electron gas\cite{giuliani_vignale}.
The flat region near the zone center ($\Gamma$) is due to self-nesting of the FS 
cylinders. This scenario is similar to the 2D electron gas where the circular
FS leads to a flat response with a sharp drop to zero at 2$k_F$. In the present
case, however, this drop is smooth, due to different sizes and irregularities in
shape of the FS cylinders. The broad peak in the $\chi$ at $M$-point is due
to the strong nesting between hole and electron like FS sheets. 
The presence of a perfect nesting, which is not realised in this 
material\cite{sharma,yaresko}, would give a diverging response at this point. 

Doping (see Fig. \ref{fig:rechix014}(a)) leads to
asymmetric change in the size of the electron and the hole FS cylinders which
manifests itself as a crater-like structure around $\Gamma$-point which 
continuously increases on going towards $M$/$X$ points in the BZ. This monotonous
increase of $\chi_{0}$ with $|q|$ indicates the predominance of inter-band 
transitions. As this inter-band nesting becomes
less  perfect upon doping (worse matching of \emph{e} and \emph{h} FS), it produces a volcano-like 
feature around the $M$-point.
This feature, of the maximum being shifted away from the 
$M$-point, is indicative of a possibility of an incommensurate spin configuration
in this material. Interestingly, this incommensuration has been recently predicted 
from two different \emph{ab-initio} calculations\cite{sharma,yaresko}.
The flatness of $\chi_{0}(\mathbf{q},0)$ in the remaining $\mathbf{q}-$space is 
consistent with the fact that all the competing magnetic structures lie within 
a small energy range, with in plane $\mathbf{q}$-vector.

\begin{figure}
\begin{tabular*}{\textwidth}{ccc}
\includegraphics[scale=0.2,clip=] {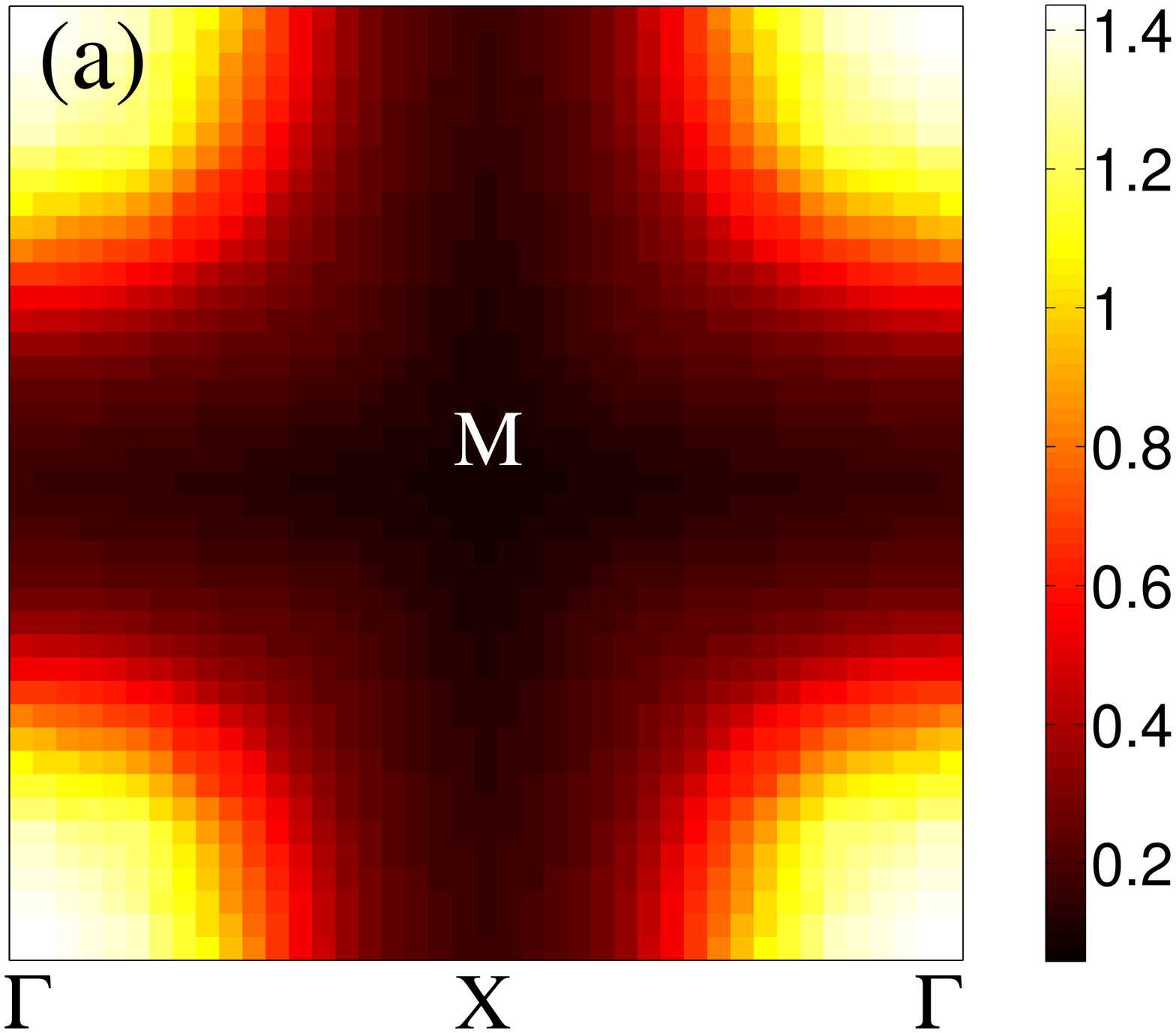} &
\includegraphics[scale=0.2,clip=] {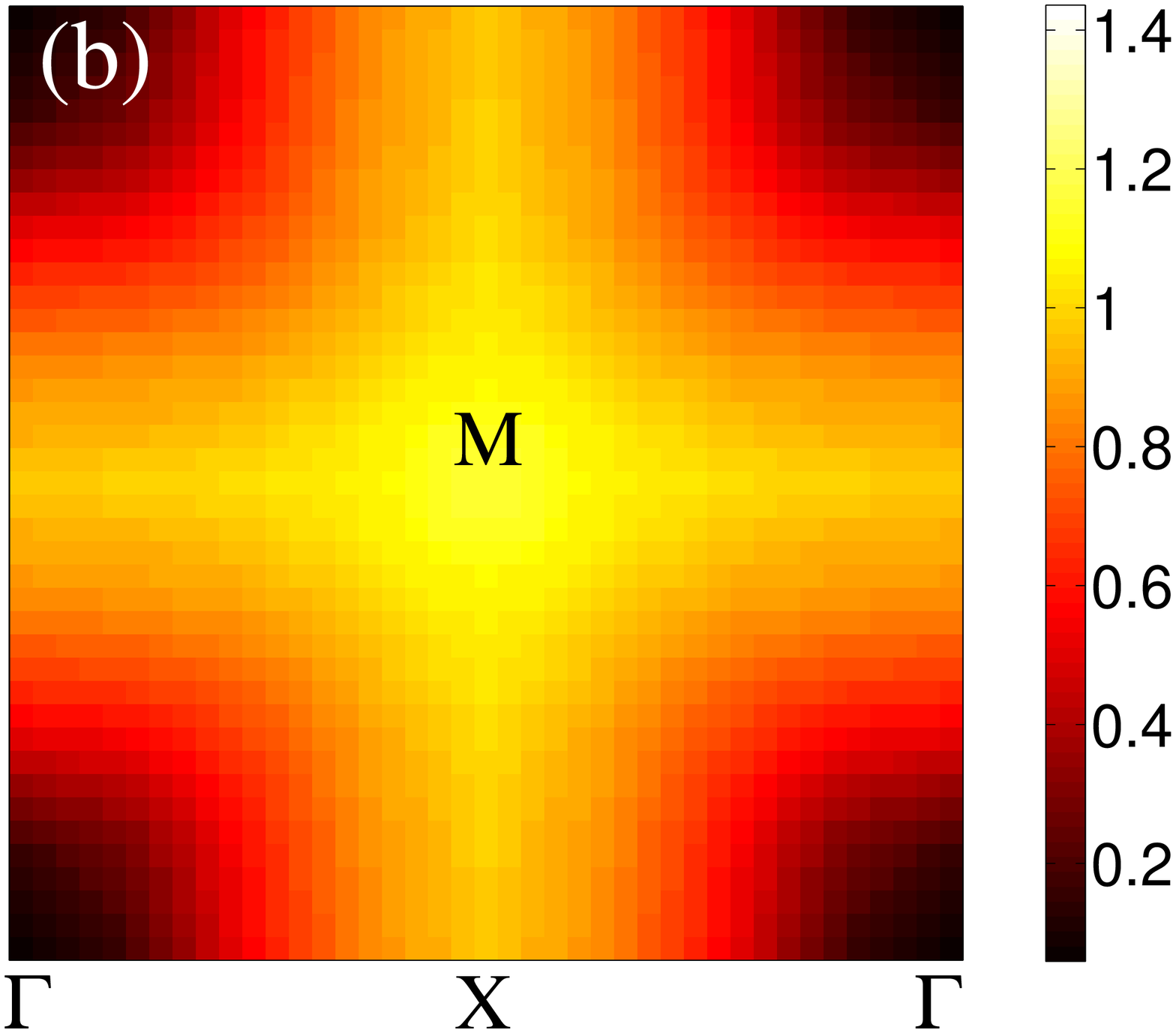} &
\includegraphics[scale=0.2,clip=] {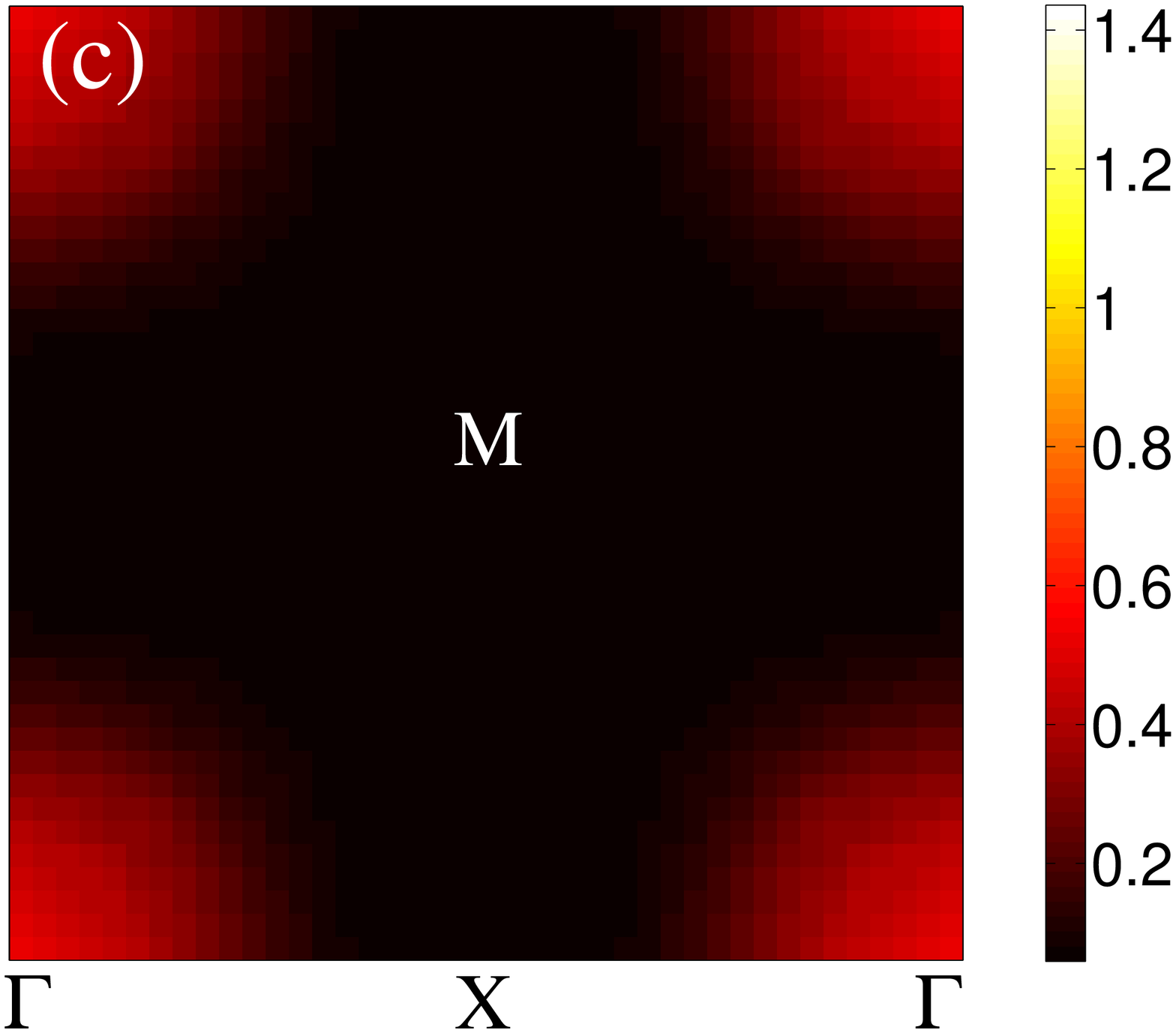} \\
\includegraphics[scale=0.2,clip=] {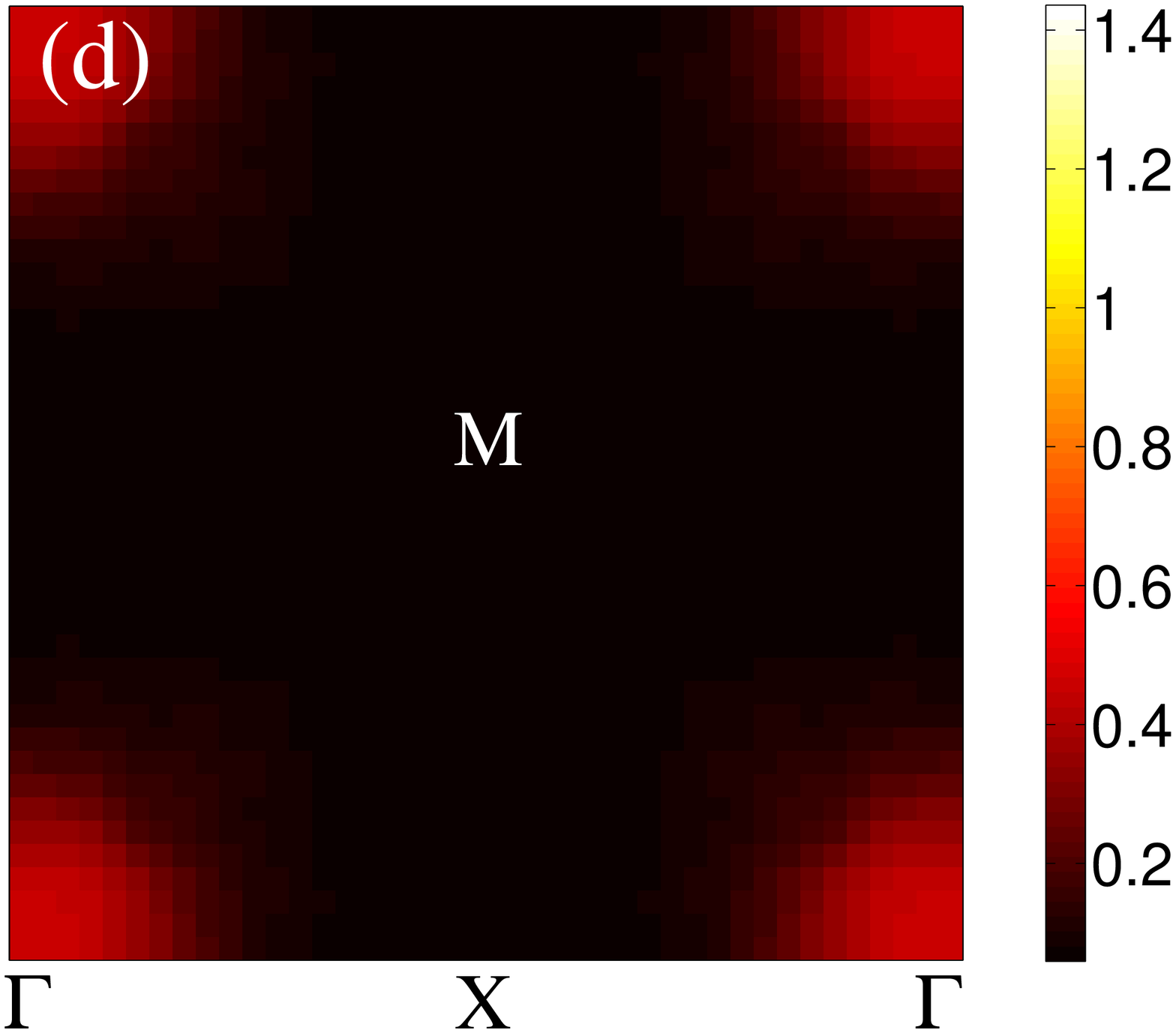} &
\includegraphics[scale=0.2,clip=] {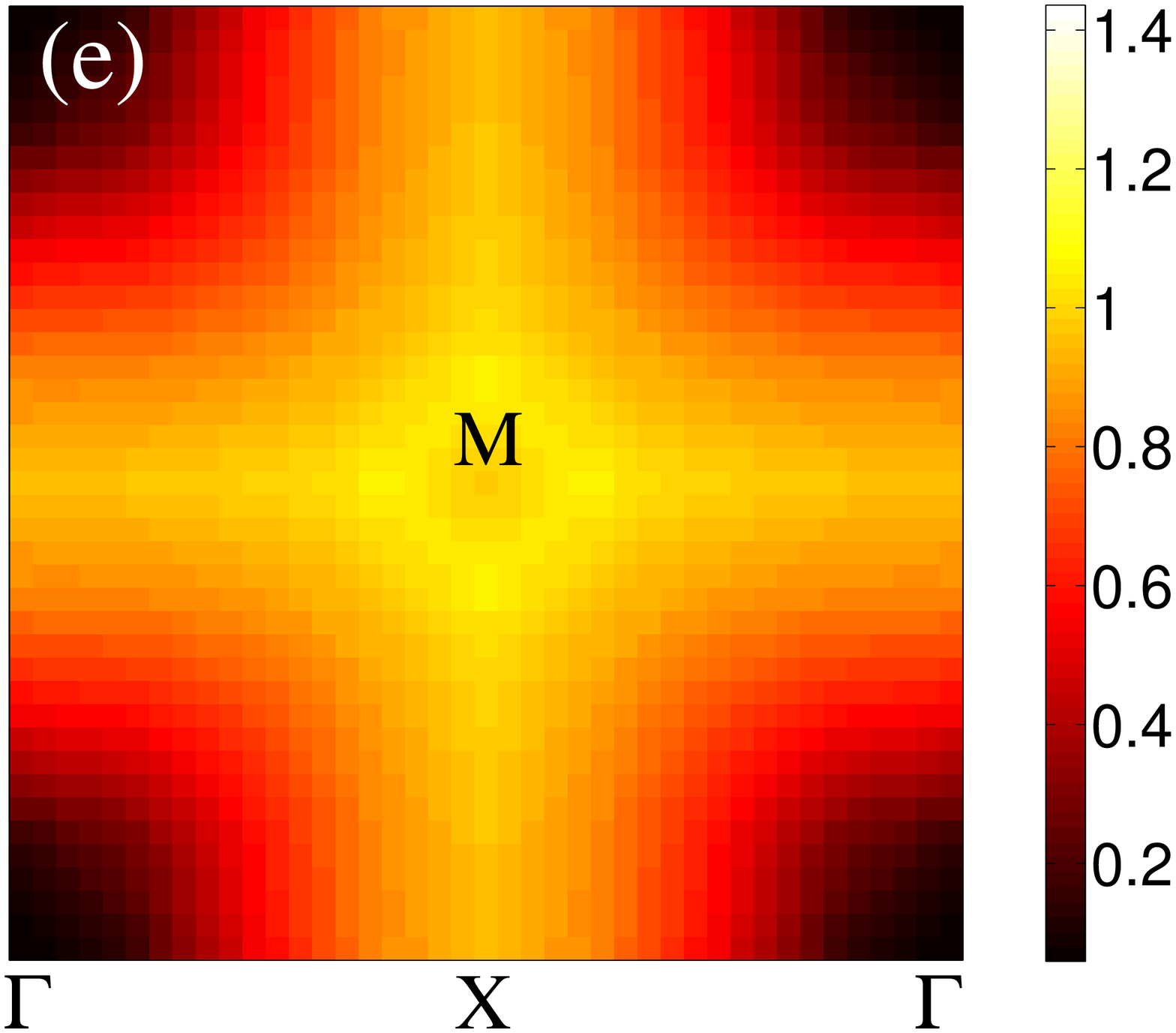} &
\includegraphics[scale=0.2,clip=] {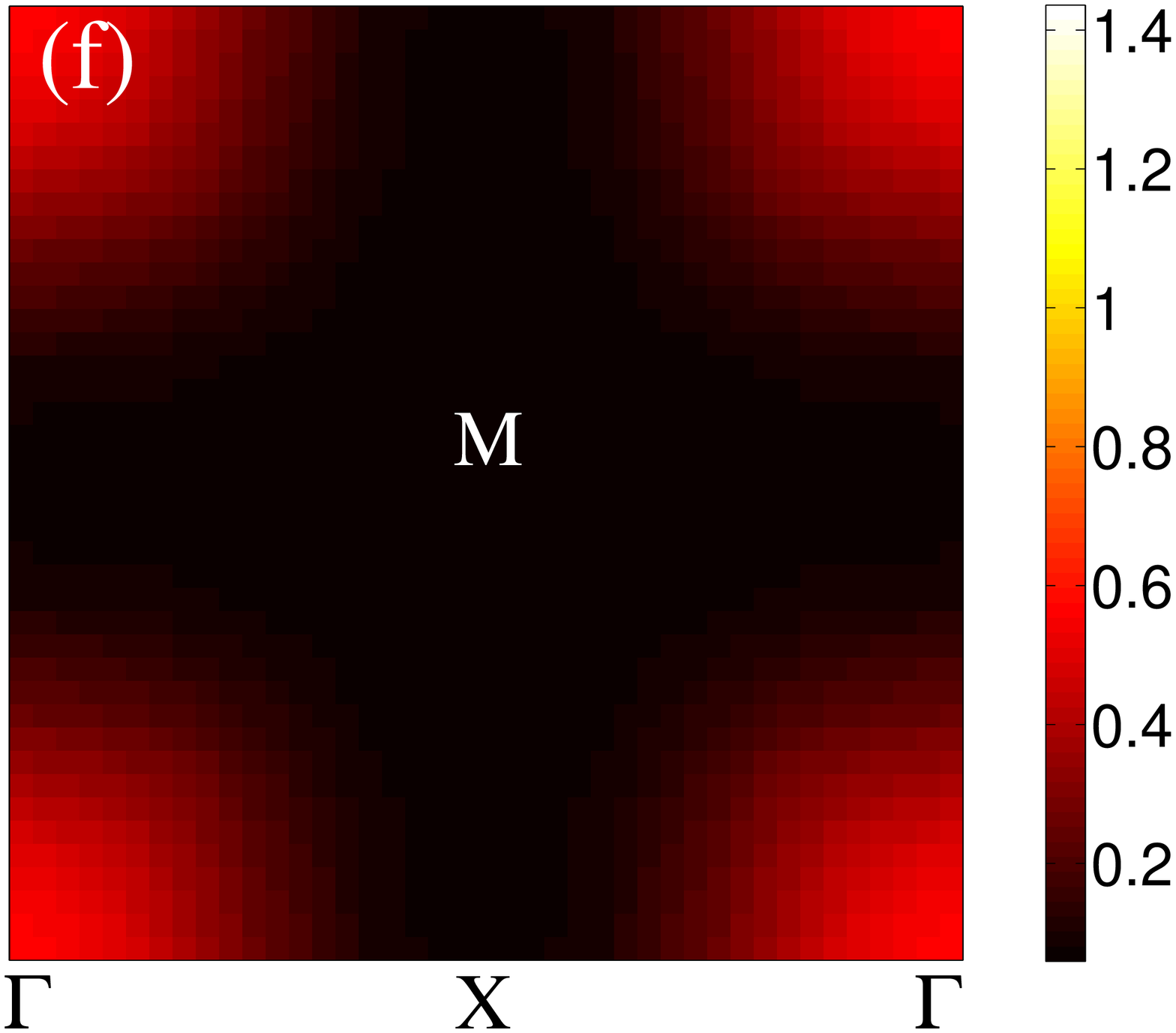}\\
\end{tabular*}
\caption{Intra- and inter-band contributions to $\chi_0(\mathbf{q},0)$,
top-panel for undoped and bottom-panel for 14\% doped LaOFeAs.
(a) and (d) are $e-e$ , 
(b) and (e) are $e-h$ and 
(c) and (f) are $h-h$ contributions.}
\label{fig:rechinnp}
\end{figure}

In order to understand the evolution of the features around the $\Gamma$ and 
$M$ points as a function of doping it is instructive to consider the contribution 
from different energy-wavevector regions to $\chi_{0}$. To 
study this we separate the contributions of different intra- and inter-band 
electronic transitions by separating the bands as follows:
the first set consists of the two bands forming \emph{h}-like  FS sheets plus all 
the occupied bands and the second set consists of the 
all bands forming electron-like FS plus all the empty bands. 
Clearly such a grouping of the bands disregards band crossings (\emph{e.g.}, those of the \emph{e} bands with filled states around $M$).
Fig. \ref{fig:rechinnp} shows static susceptibility resolved according to 
this criteria. Looking at various contributions around the $\Gamma$ point it
becomes clear that in both doped and undoped cases the low momentum susceptibility is
dominated by the intra-band contribution ($e-e$ and $h-h$). Doping leads to
a strong reduction in the $h-h$ contribution, which is expected since the effect 
of doping is to shrink the hole FS sheets around the $\Gamma$ point. On the other hand 
around the $M$ point and in the rest of the plane the inter-band $e-h$ contributions 
dominate. Upon doping this inter-band contribution does not change substantially
except in the very vicinity of the $M$ point where a volcano like feature is formed. 
This can be easily understood as a consequence of the change of shape of the FS 
upon doping; the asymmetry between the \emph{e} and \emph{h} sheets becomes stronger 
resulting in a reduction of the strength of the nesting. 

\begin{figure}[htbp]
\begin{center}
\includegraphics[scale=0.165]{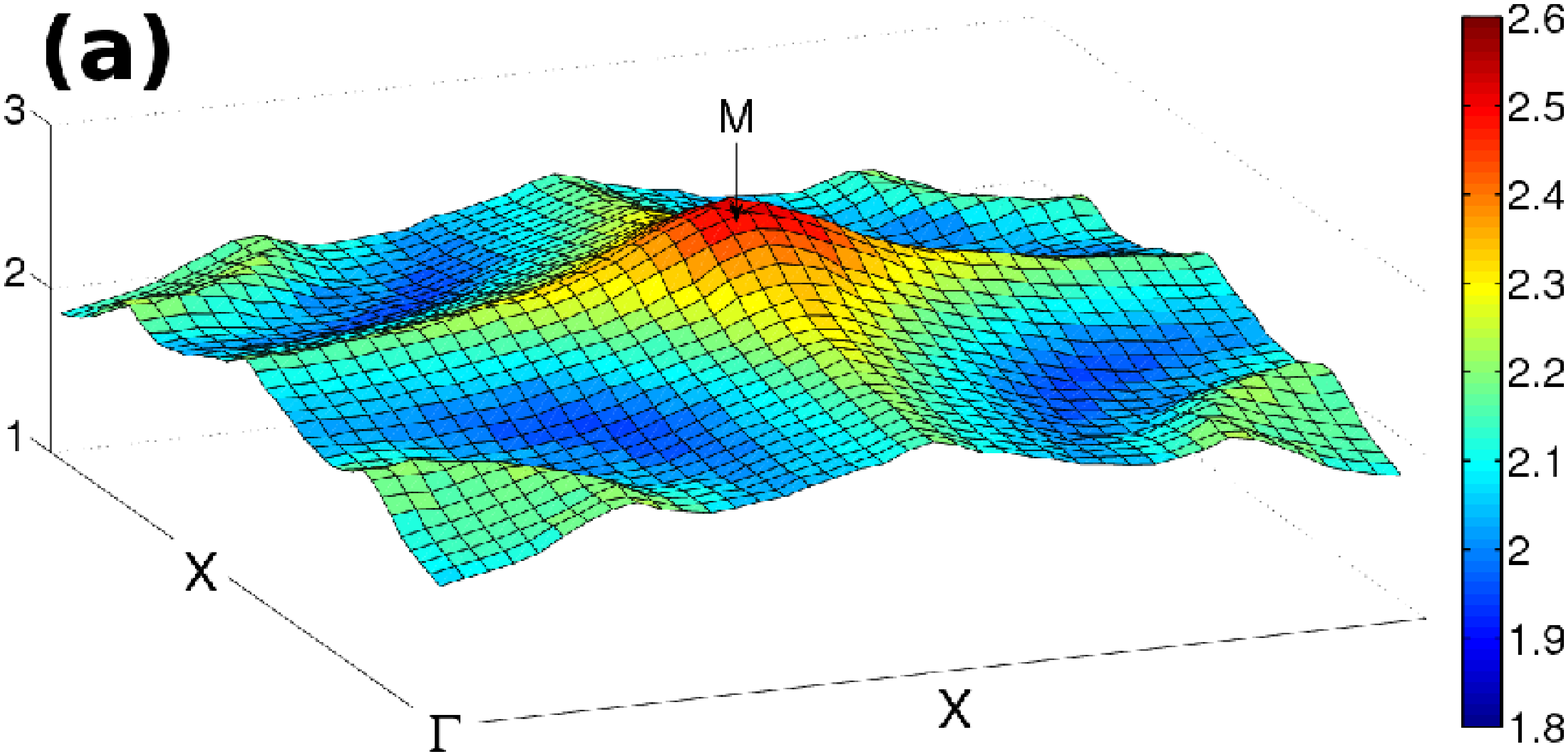}
\includegraphics[scale=0.165] {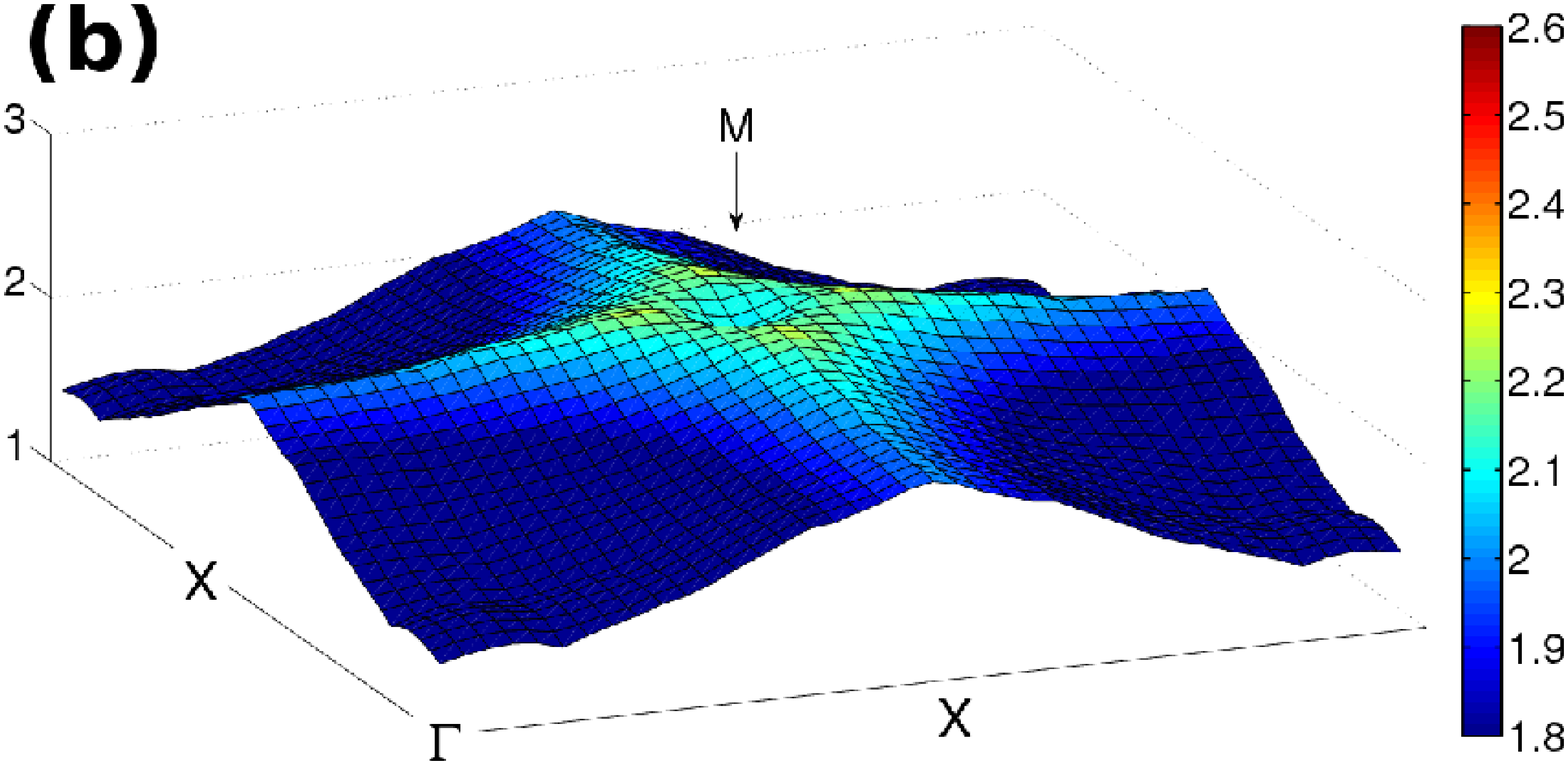}\linebreak
\includegraphics[scale=0.165] {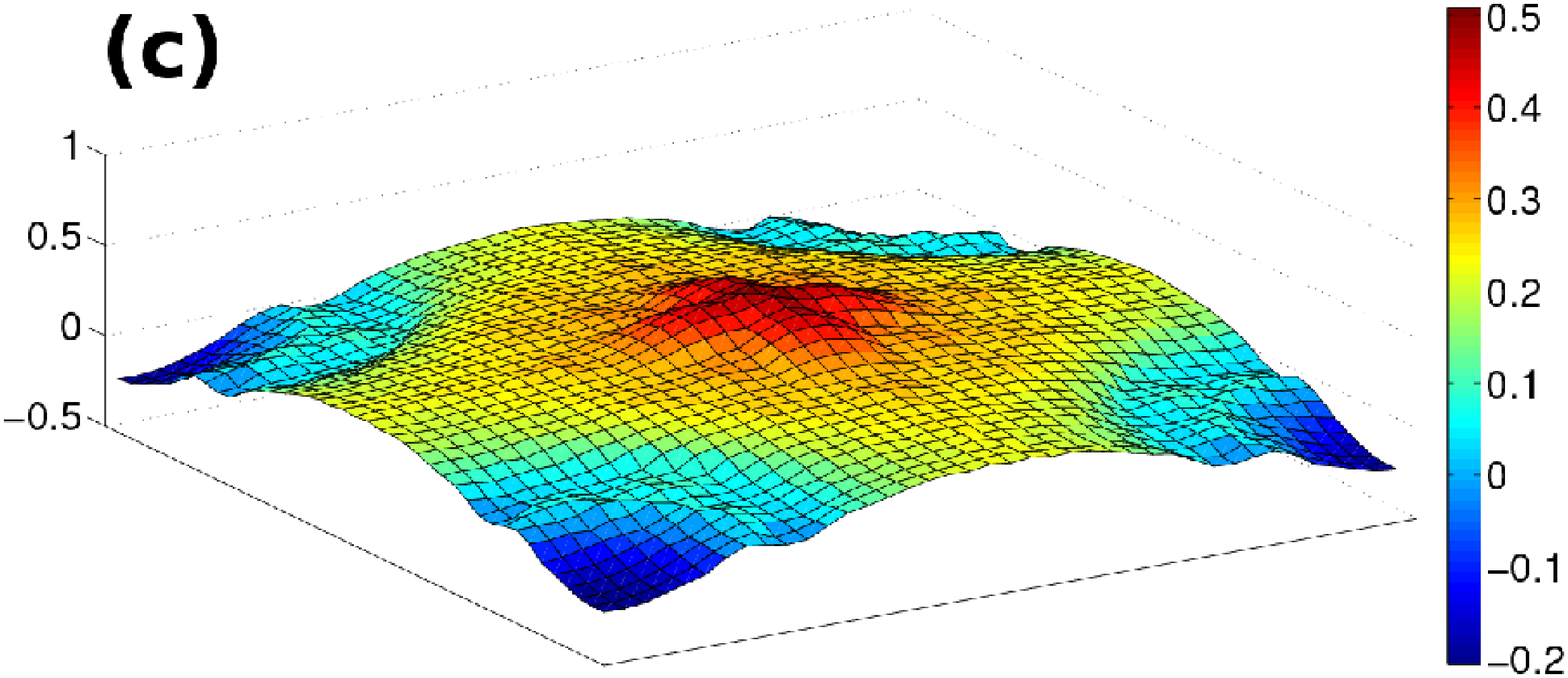}
\includegraphics[scale=0.165]{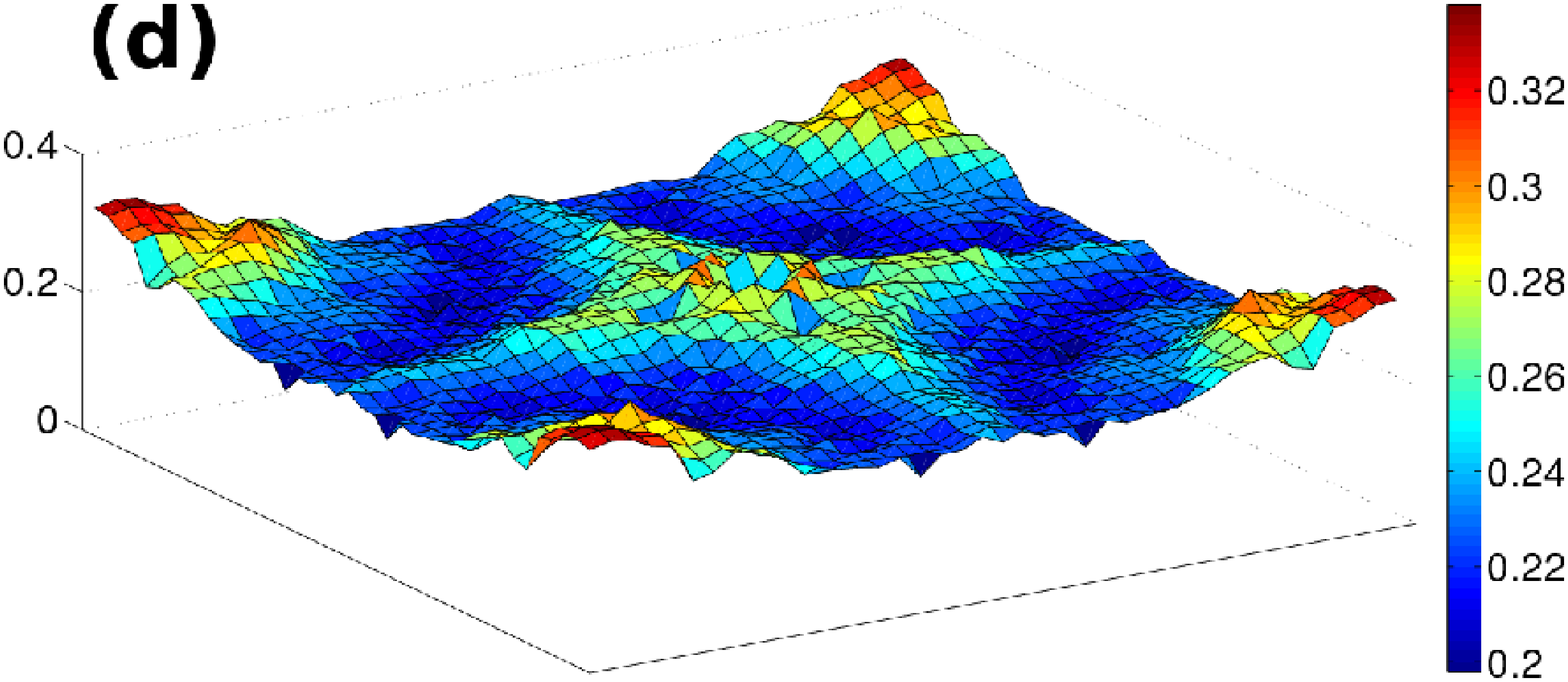}
\caption{Real part of $\chi(\mathbf{q\mathrm{,0)}}$ on the the basal plane (i.e. $q_{z}=0$),
calculated using the experimental position of the As atom. 
Top-panel: (a) for the undoped case and (b) for the 14\% doped case.
Bottom-panel: (c) difference between (a) and same calculated at the 
theoretically optimized position of the As atom and (d) difference between (b) 
and same calculated at the theoretically optimized position of the As atom.} 
\label{fig:chi_position}
\end{center}
\end{figure}

In order to investigate how the response of LaO$_{1-x}$F$_{x}$FeAs is affected 
by the change in atomic positions, we also calculate the $\chi_{0}^{(exp)}$, at the 
experimental position of the As atom. In Fig.~\ref{fig:chi_position} we plot 
$\chi_{0}^{(exp)}$ (in the basal plane) as a function of doping, and its change 
relative to the $\chi_{0}$ calculated at the optimized position of the As atom.
We see a sizeable difference between the two  at zero doping: using experimental 
positions, the peak at the nesting vector is much more evident and broader, 
while the peak around $\Gamma$ is depressed. We can relate this change to the 
differences in the electronic structure, the 3D bands crossing $E_{F}$ around 
the $Z$ point become 2D, but also the radius of the FS cylinders is larger in 
the optimized structure making the nesting between {\emph e}-like and 
{\emph h}-like sheets less efficient. In the doped system, on the other hand, 
this band is completely full in both atomic configurations as a consequence of 
which in the doped case, the difference induced by change in atomic positions 
is not as significant.

On comparing the static susceptibility calculated in the present work
with the previous results of Mazin et al. \cite{Mazin} one
notices that the height of the peak around the $M$ point is significantly suppressed 
on inclusion of the matrix elements. This is a consequence of the fact that the
states near $E_{F}$  contribute to ${\rm Re} \chi_{0}$ selectively, weighted by their 
overlap integral and not by unity. 

It is believed that the interacting susceptibility is a key quantity in determining the 
effective electron-electron interactions leading to superconductivity. Our results 
show  that the inclusion of the matrix elements influences its ${\bf q}$ 
dependence and in particular the relative  height of the peaks; 
Although our calculations only refer to the independent-electron susceptibility , 
these conclusions are likely to hold also for the interacting $\chi$. Therefore, it is important to 
include the matrix elements in order to use $\chi$ for assessing the validity of 
models for the superconducting  mechanism (\emph{e.g.} the $s_\pm$ wave).

\subsection{Dynamical susceptibility }
\begin{figure}
\includegraphics[scale=0.4]{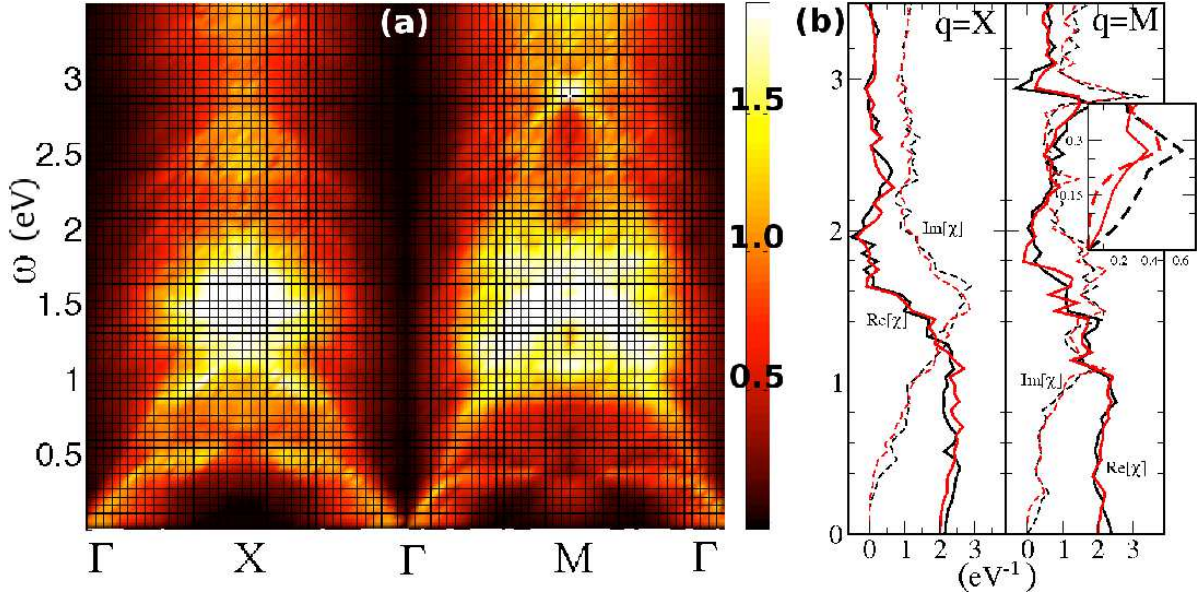}
\caption{(color online)
(a) Imaginary part of $\chi_{0}(\mathbf{q},\omega)$ (eV), 
for undoped LaOFeAs, as a function of $\mathbf{q}$ and $\omega$. 
Results are plotted along $\Gamma-X-\Gamma- M-\Gamma$ directions.
(b) Plot of the real and imaginary parts of $\chi_{0}$ as a function of $\omega$ for 
$\mathbf{q}$ corresponding to the $X$ and $M$ points of the BZ.  Black and red 
refer to $x=0$ and $x=0.14$ respectively. The inset of (b) shows the low frequency part of 
$\chi_{0}$ at $M$ for $x=0$ and $x=0.14$ (black and red dashed respectively), and at one $q$-point
point close to $M$ corresponding to the edge of the volcano-like structure of $\rm{Re} \chi_{0}$ (thin red line).
}
\label{fig:imchi}
\end{figure}

An overview of the dynamical response of the system is given in Fig.~\ref{fig:imchi},
where we plot $\rm {Im} \chi_{0}$, the imaginary part of $\chi_{0}\left(\mathbf{q\mathrm{,\omega}}\right)$,
as a function of frequency and momentum, along the $\Gamma-M-\Gamma$ and $\Gamma-X-\Gamma$ lines of the Brillouin zone, for $x=0$. 
The most prominent structure is a broad peak located at frequencies between 
1 and 2 eV, which behaves in a similar manner along the  $\Gamma-M-\Gamma$ and 
$\Gamma-X-\Gamma$ directions. Near the $M$ point we also see a sharp and 
intense peak located at about 3 eV. At this point it is also worth mentioning 
that the flat La $4f$ bands visible at
around  3 eV in Fig.~\ref{fig:bande} do not result into any evident structure 
in the ${\rm Im}\chi_{0}$ plot, indicating the small hybridization 
(thereby small matrix element) of these states with Fe and As states.

At small values of $\mathbf{q}$ and at low
frequency, the value of $\rm{Im} \chi_{0}$ grows linearly as a function of the 
frequency and then drops rapidly to zero (resembling linear response susceptibility for the non-interacting electron gas). 
From Fig.~\ref{fig:imchi}(a), we see that this low
frequency feature has two components; the position of the  first
peak grows with {\bf q}, saturating at a frequency of $\sim$0.4 eV, while
the second one grows up to  $\sim$ 1eV finally merging into the main broad peak.
This behaviour is the same along the  $\Gamma-M-\Gamma$ and $\Gamma-X-\Gamma$ 
lines, except for the presence of an extra  low frequency  structure ($\omega \leq 0.3$ eV) visible around the 
$M$-point.
A detailed view of $\rm{Im} \chi_{0}(\mathbf{q},\omega)$ and the behaviour of $\rm{Re} \chi_{0}(\mathbf{q},\omega)$
is given in Fig.~\ref{fig:imchi}(b) at $X$ and $M$, both for the undoped and for the doped systems. 
Even though doping does not significantly change the general shape of both real and imaginary parts of $\chi_{0}$, 
in the low frequency region, we notice a different behaviour of  $\rm{Im} \chi_{0}(\mathbf{q},\omega)$ at $M$ (see inset of Fig. 7b). 
For $x=0$ the Im$\chi_0$ grows linearly as a function of $\omega$, while it only starts to grow at a finite value of $\omega$ for $x=0.14$. 
The reason for this is closely tied to the nesting function; 
An highly nested FS gives rise to linear behaviour for $x=0$.
While  in the doped case, deterioration  of FS nesting at $M$, only allows finite energy excitations, leading to a finite starting value for the $\rm{Im} \chi_{0}(\mathbf{q},\omega)$.
Interestingly, if we move slightly away from $M$ (on the edges of the volcano structure discussed above), we recover a linear trend starting at $\omega=0$.

\begin{figure}
\includegraphics[scale=0.18]{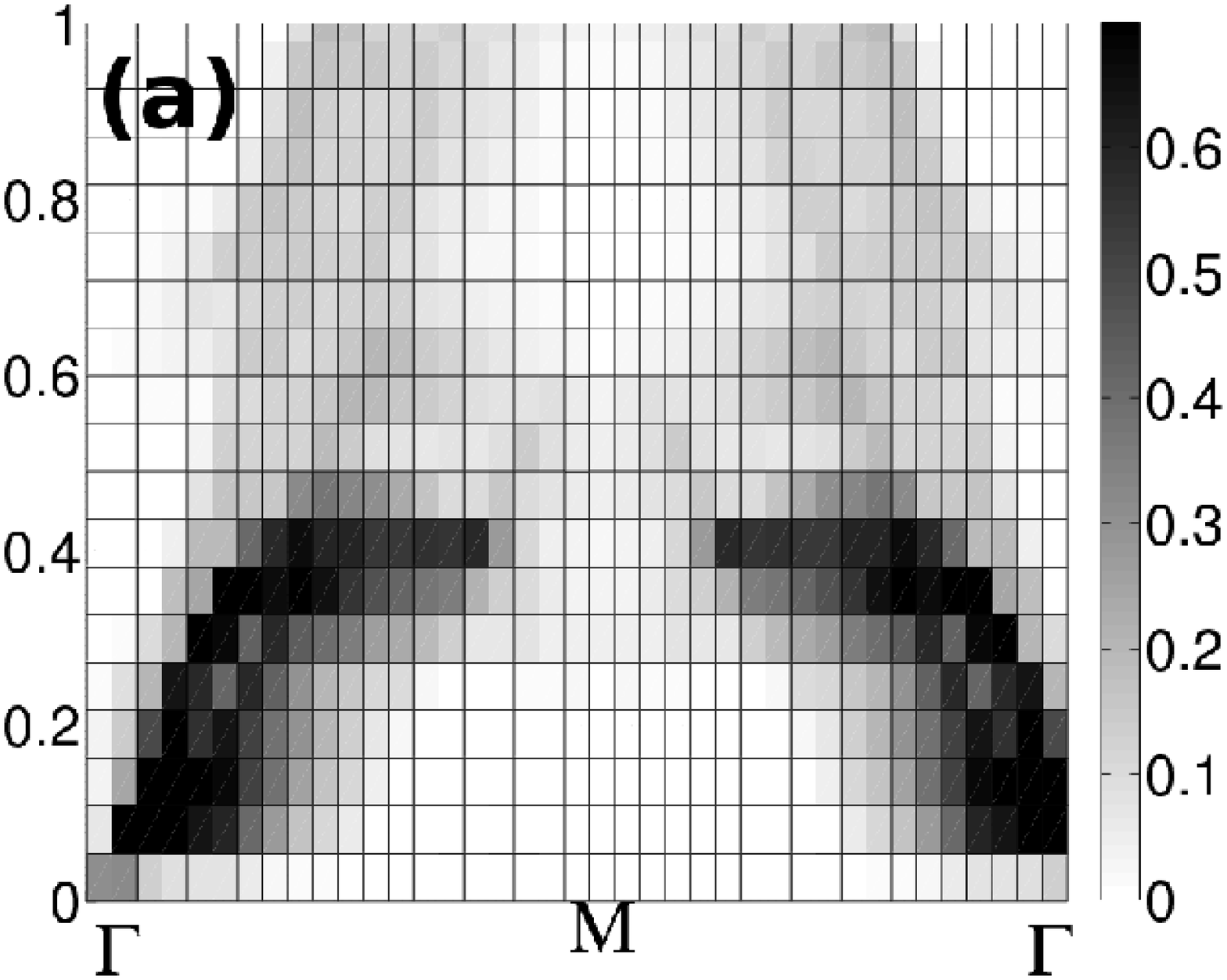}
\includegraphics[scale=0.18]{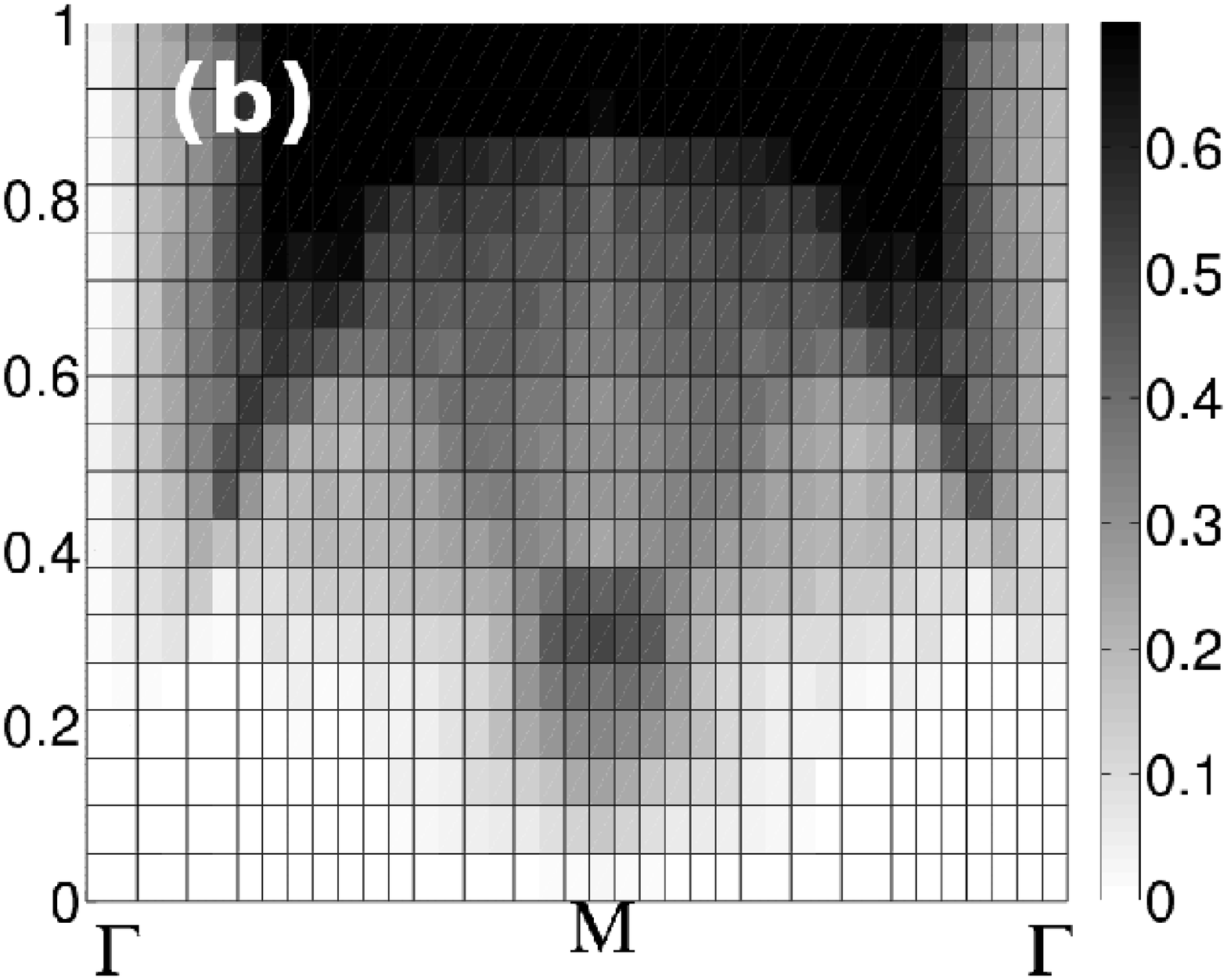}
\includegraphics[scale=0.18]{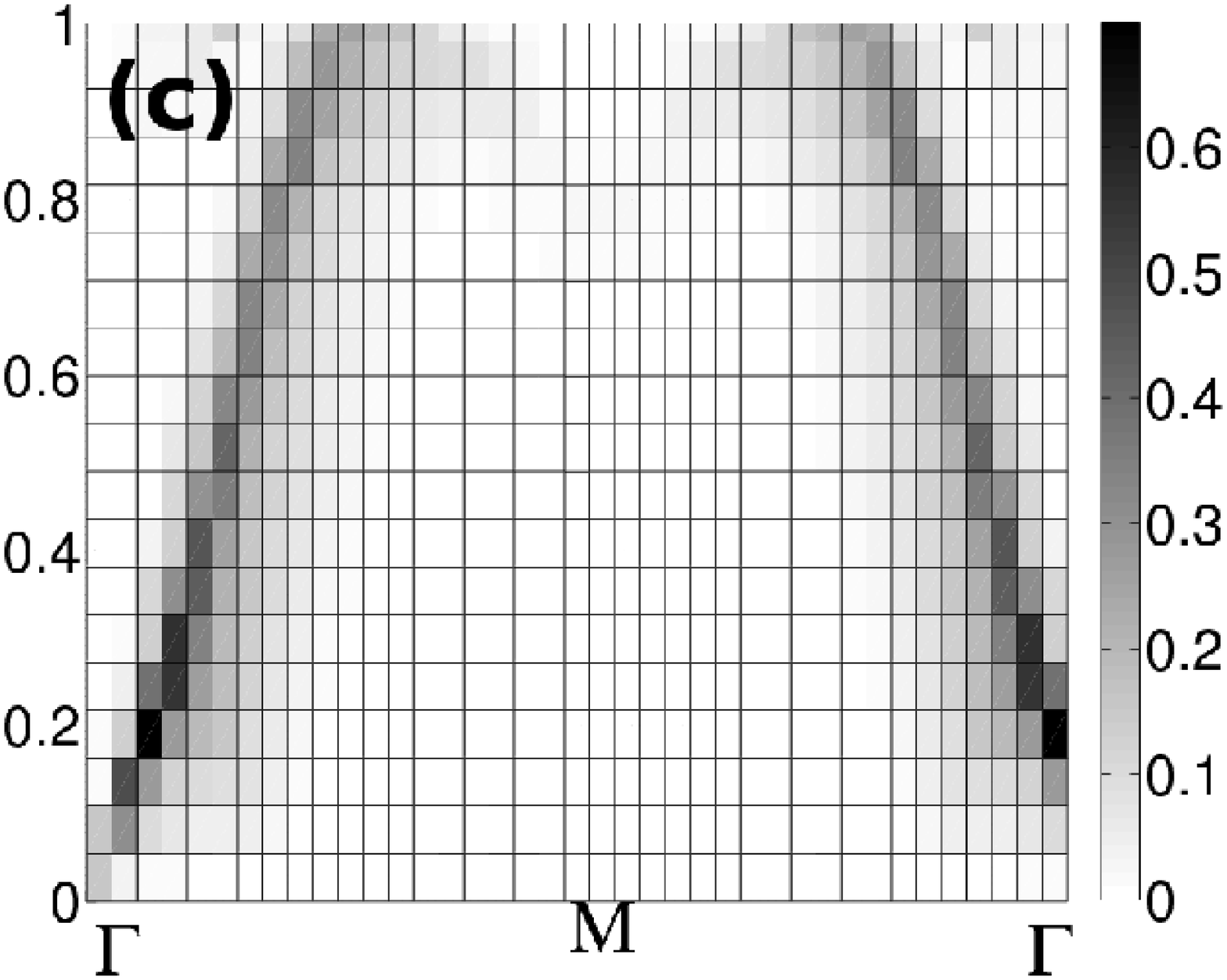}
\caption{ Imaginary part of $\chi_{0}(\mathbf{q},\omega)$ (eV) decomposed into 
the intra- and inter-band contributions. Results are plotted along  
$\Gamma- M-\Gamma$ direction. (a) $h-h$, (b) $e-h$ and (c) $e-e$ contributions.}
\label{fig:im2x2}
\end{figure}

The band decomposition of ${\rm Im}\chi_{0}\left(\mathbf{q\mathrm{,\omega}}\right)$,
reported in Fig.~\ref{fig:im2x2} along the $\Gamma-M-\Gamma$ line,
shows that the inter-band ($e-h$) contributions dominate the high frequency 
part, and are also responsible for the low frequency peak around $M$. 
The two dispersive peaks at low frequency discussed above originate from 
the intra-band transitions ($e-e$ and $h-h$); in particular, the relatively high 
frequency branch comes mainly from electronic bands, while the low frequency one from 
hole bands. Fig.~\ref{fig:im2x2}  clearly shows that the $h-h$ and $e-e$ 
contributions are quite asymmetric; this asymmetry is clearly a consequence of 
the richness of the  electronic structure of LaOFeAs near $E_{F}$.

\section{Conclusion}
\label{concl}
In summary, we report detailed calculations of  the
independent electron susceptibility of non-magnetic LaO$_{1-x}$F$_{x}$FeAs 
as a function of doping and of the  atomic positions within the unit cell. 
Our results  are based on accurate electronic structure 
calculations within density functional theory, and  include  matrix elements 
from full potential linearized augmented plane wave method. We account properly for
Fermi surface related features through an accurate sampling of the Brillouin zone.

The static susceptibility is peaked around the zone center and at the nesting vector
$\mathbf{q}_{N}$  ($M$ point), due to intra-band ($e-e$ and $h-h$) and 
inter-band transitions respectively, and is 
consistent with the observed stripe
AFM ordering. 
However, the peak at $M$-point is not as pronounced as 
reported in calculations with approximate or no matrix elements.  
Upon doping, the peak at $M$ evolves into a volcano-like structure consistent with the 
incommensurate  magnetic spiral state predicted by first principle calculations.
The intra- and inter-band analysis of the contributions 
to $\chi_{0}(q,0)$ shows an  \emph{e} versus \emph{h} asymmetry which 
may relate to the multigap character suggested by experiments.
Our results could serve as a first step towards the definition of an \emph{ab-initio} 
effective electron-electron interaction, necessary to obtain the pairing potential in pnictides.

\begin{acknowledgments}
We acknowledge computational support by COSMOLAB consortium (Cagliari, Italy).
by INFM-CNR through a supercomputing grant at Cineca (Bologna, Italy),
by the Deutsche Forschungsgemeinschaft and by
NANOQUANTA Network of Excellence.
F.B. acknowledge CASPUR for support by the HPC grant 2009. 
M.M. acknowledge support from the Regione Sardegna, through the "Master and Back" program.
\end{acknowledgments}

\end{document}